\documentclass[a4paper]{article}
\pdfoutput=1
\usepackage{jheppub}
\usepackage{amsmath}
\allowdisplaybreaks[4]
\usepackage[utf8]{inputenc}
\usepackage{float}
\usepackage{amssymb}
\usepackage{graphicx,color}
\usepackage{amsmath,amsthm}
\usepackage{diagbox}
\usepackage{mathrsfs}
\usepackage{float}
\usepackage{bm}
\usepackage{mwe}
\usepackage[toc,page]{appendix}
\usepackage{url}
\usepackage{varioref}

\def\f{\frac}

\newcommand{\ket}[1]{| #1 \rangle}

\newtheorem{Theorem}{Theorem}[section]

\newcommand{\Z}{\mathbb{Z}}

\newcommand{\C}{\mathbb{C}}

\newcommand{\half}{\frac{1}{2}}
\newcommand{\double}[2]{\hspace{0.1em} #1 \hspace{#2} #1 \hspace{0.1em}}

\newcommand{\ident}{\double{1}{-0.35em}}

\newcommand{\ccirc}{\kern0.2ex\vcenter{\hbox{$\scriptstyle\circ$}}\kern0.2ex}

\newcommand{\e}{\mathrm{e}}

\newcommand{\Slc}{\mathrm{SL}(2,\mathbb{C})}

\newcommand{\Su}{\mathrm{SU}(2)}

\def\be{\begin{eqnarray}}
\def\ee{\end{eqnarray}}

\def\mb{\left(\begin{matrix}}
	\def\me{\end{matrix}\right)}

\newcommand{\cl}{\mathcal L}

\renewcommand{\a}{\alpha}

\newcommand{\g}{\gamma}
\newcommand{\G}{\Gamma}
\renewcommand{\d}{\delta}

\newcommand{\sig}{\sigma}

\renewcommand{\l}{\lambda}

\newcommand{\z}{\zeta}

\newcommand{\rmd}{\mathrm d}

\newcommand{\lt}{\left}
\newcommand{\rt}{\right}

\title{Numerical computations of next-to-leading order corrections in spinfoam large-$j$ asymptotics}

\author[1,2]{Muxin Han}  

\affiliation[1]{Department of Physics, Florida Atlantic University, 777 Glades Road, Boca Raton, FL 33431-0991, USA}

\affiliation[2]{Institut f\"ur Quantengravitation, Universit\"at Erlangen-N\"urnberg, Staudtstr. 7/B2, 91058 Erlangen, Germany}
\author[3,4]{\ Zichang Huang}
\affiliation[3]{Department of Physics, Center for Field Theory and Particle Physics,
	and Institute for Nano- electronic devices and Quantum computing, Fudan University, Shanghai 200433, China}
\affiliation[4]{State Key Laboratory of Surface Physics, Fudan University, Shanghai 200433, China}

\author[2]{\ Hongguang Liu}  
\author[1]{\ Dongxue Qu}

\emailAdd{dqu2017(At)fau.edu}

\abstract{We numerically study the next-to-leading order corrections of the Lorentzian Engle-Pereira-Rovelli-Livine (EPRL) 4-simplex amplitude in the large-$j$ expansions. We perform large-$j$ expansions of Lorentzian EPRL 4-simplex amplitudes with two different types of boundary states, the coherent intertwiners and the coherent spin-network, and numerically compute the leading-order and the next-to-leading order $O(1/j)$ contributions of these amplitudes. We also study the dependences of these $O(1/j)$ corrections on the Barbero-Immirzi parameter $\gamma$. We show that they, as functions of $\gamma$, stabilize to finite real constants as $\gamma\to\infty$. Lastly, we obtain the quantum corrections to the Regge action because of the $O(1/j)$ contribution to the spinfoam amplitude.}

\begin{document}

\maketitle
\section{Introduction}

\par Loop Quantum Gravity (LQG) is a candidate for background-independent and non-perturbative quantum theory of gravity \cite{Thiemann:2007pyv,review1,Han:2005km}. The spinfoam model is a covariant approach to Loop Quantum Gravity, and it provides LQG transition amplitudes, the spinfoam amplitude, as a sum-over-history of quantum geometries \cite{rovelli2014covariant,Perez2012}. Because of the simplicity and semi-classical behavior \cite{Rovelli:2010vv,Barrett:2009mw,Han:2011re} of the Lorentzian Engle-Pereira-Rovelli-Livine (EPRL) model \cite{Engle:2007wy},  it is one of the most successful spinfoam models. In the Lorentzian EPRL model, the spinfoam amplitude can be described by a path integral representation that is employed in studying the large-$j$ asymptotic behavior. This asymptotic behavior is related to the Regge action of the classical discrete gravity \cite{Conrady:2008ea, Han:2013gna}. Computing spinfoam amplitudes is central in developing the spinfoam formulation of LQG, especially from the perspective of extracting quantum corrections to the classical gravity. Existing studies on the Lorentzian EPRL model mainly focus on the leading order contribution in the large-$j$ asymptotics, and leave the higher order corrections unexplored. Higher order corrections in the large-$j$ expansion are expected to relate to the quantum-gravity effects in LQG, while the leading-order terms relate to the semi-classical limit.

The purpose of this paper is to study the next-to-leading corrections in the large-$j$ expansion of the Lorentzian EPRL 4-simplex amplitude with two types of boundary states which are coherent intertwiners and coherent spin-networks. Here we introduce the main results of this paper. We consider the same Lorentzian non-degenerate 4-simplex geometry and boundary data as \cite{Dona:2019dkf} and construct spinfoam critical points of the EPRL amplitude. For the coherent intertwiners as the boundary state, there are two critical points (of opposite 4-simplex orientations). Following the asymptotic expansion (H\"ormander's theorem 7.7.5 in \cite{Hormander}), we perform large $j$ asymptotic expansion of the 4-simplex amplitude at both critical points, and numerically compute both the leading-order and the next-to-leading order corrections. If we scale spins by $j_{f}\to\l j_f$ for all boundary triangles $f$, the expansion in $\l$ is represented as below 
\be
A_{v}^{(\pm)} = C^{(\pm)}(\gamma) \cdot\left[1+\frac{\kappa^{(\pm)}(\gamma)}{\lambda}+O\lt(\frac{1}{\l^2}\rt)\right]\label{Amplitude_1}
\ee  
where $C^{(\pm)}, \kappa^{(\pm)}$ depending on the value of $\g$ are computed numerically in this work. $C^{(\pm)}$ is identical to the leading-order asymptotics given by Barrett et al \cite{Barrett:2009mw}. The evaluation of the next-to-leading order coefficient $\kappa^{(\pm)}(\g)$ is one of main interests in this work. It turns out that $\kappa^{(+)}(\g)=\overline{\kappa^{(-)}(\g)}$. As an example, at $\g=0.1$, $|\kappa^{(\pm)}(0.1)|\simeq 3.14$ and the 4-simplex amplitude $A_v=A_{v}^{(+)}+A_{v}^{(-)}$ is given by\footnote{The next-to-leading order gives a sine function similar to the expansion of 6j symbol \cite{Bonzom:2008xd}. }
\be
A_v&=&\left(1+\frac{1}{4 \lambda }\right)^6\left(1+\frac{1}{10 \lambda }\right)^4\frac{3.55 \times 10^{-13}}{\lambda^{12}}\,e^{4.59 \lambda i}\nonumber\\
&&\left[\cos (0.106+0.01 \lambda)+\frac{3.14}{\lambda} \sin (-1.27+0.01 \lambda)+O\lt(\frac{1}{\l^2}\rt)\right],\label{Avresult000}
\ee 
where $S_{Regge}=0.01 \lambda$ in the terms of cosine and sine is the Regge action of the geometrical 4-simplex. The next-to-leading order corrections have to be sufficiently small in order to validate the semiclassical approximation of $A_v$ with the leading-order terms as \cite{Barrett:2009mw}. By the above result, for example, when $\l=30$, the magnitude of the second term in Eq.(\ref{Amplitude_1}), $|{\kappa^{(\pm)}(0.1)}/{\lambda}|\simeq 0.1$, is about 10\% of the leading-order terms. We can conclude that approximating the amplitude $A_{v}^{(\pm)}$ solely by the leading order term $C^{(\pm)}(\g)$ leads to an error about 10\% in the case of $\g=0.1$ and $\l=30$. The similar behaviors are supported by several numerical examples with different boundary geometries.   

This conclusion may become different when we impose the different boundary state. We consider the boundary state to be the coherent spin-networks in Section \ref{spinnetwork}. In this case, the EPRL amplitude $A'_v$ contains summing over $j$. The boundary coherent spin-networks determine one critical point of the amplitude while eliminating the others. The asymptotic expansion gives
\be
A'_{v} = C'(\gamma) \cdot\left[1+\frac{\kappa'(\gamma)}{\lambda}+O\lt(\frac{1}{\l^2}\rt)\right].
\ee 
We find that in certain example of 4-simplex geometry and boundary data, at $\g=0.1$, the next-to-leading order coefficient gives $|\kappa'(0.1)|\simeq 40.67$. When $\l=30$, $|{\kappa'(0.1)}/{\lambda}|\simeq 1.36$ is even larger than the leading-order term. Clearly, the semiclassical approximation of $A'_v$ is invalid at $\l=30$, and a much larger $\l$ is needed. For instance, when $\l\geq 300$, $|{\kappa'(0.1)}/{\lambda}|$ is bounded by about 13\% of the leading-order term. We suggest a much safer zone to be $\l\geq 3000$ for $A'_v$ ($\l\geq 300$ for $A_v$) where the next-to-leading order correction is about $1\%$ of the leading-order term. However, we find this increase of allowed $\l$ when $\g$ is small is not universal, it only happens in certain examples.


Moreover, we numerically study the dependences of $\kappa^{(\pm)}$ and $\kappa'$ on $\g$ in several examples with different boundary geometries. Numerical results support that they stabilize to real constants asymptotically as $\g\to\infty$.


Main computations in this work are carried out by Mathematica. Mathematica codes for constructing critical points and computing large-$j$ expansion can be found in \cite{qudx.org}. Although our computation fixes the 4-simplex boundary data, the codes can be easily adapted to other boundary data.  
\par In addition, $A_v$ in Eq.(\ref{Avresult000}) can be rewritten (up to an overall phase) as
\be
A_v&\simeq&\frac{1}{\lambda^{12}}  \lt( e^{S^{(+)}_{eff}}+e^{S^{(-)}_{eff}}\rt).\label{result2}
\ee 
where 
\be
S^{(\pm)}_{eff}=\pm i\lt(0.01\lambda+0.106-\frac{0.601}{\lambda}\rt)-28.6667 - \frac{1.182}{\l}
\ee
can be viewed as the ``quantum effective action'' with quantum corrections to the Regge action $S_{Regge}=0.01\l$.

Here are some other works on numerical analysis of spinfoam models from different perspectives: \cite{Dona:2018nev,Dona:2019dkf} numerically compute  the EPRL amplitude in the spin-intertwiner representation, by decomposing Clebsch-Gordan coefficients of SL(2,C) in terms of those of SU(2). \cite{Bahr:2016hwc,Bahr:2018gwf} numerically compute symmetry-restricted spinfoam models and their renormalization.

This paper is organized as follows:  Section \ref{spinfoam} is a brief review of the EPRL 4-simplex amplitude. Section \ref{bd} explains the boundary data and the construction of critical points. New results of this paper start from Section \ref{intertwiner}, where we expand the amplitude with the coherent intertwiner boundary state and numerically compute  both the leading-order terms and the next-to-leading order corrections for various values of $\g$. Section \ref{spinnetwork}, we study the EPRL amplitude with the boundary coherent spin-network, and numerically compute both the leading-order terms and next-to-leading-order corrections for various values of $\g$.

\section{EPRL 4-simplex amplitude}\label{spinfoam}

\begin{figure}[t]	
\centering
\includegraphics[scale=0.5]{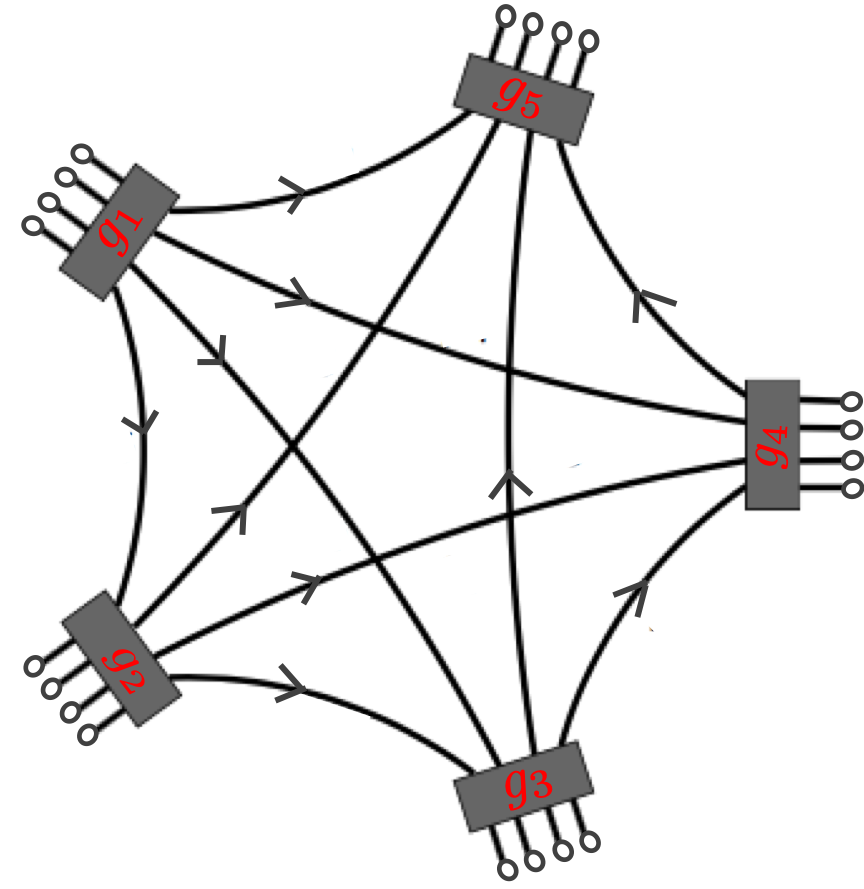}
\caption{The graphical illustration of the 4-simplex amplitude: Five black boxes correspond to boundary tetrahedra carrying $g_{a} \in \Slc (a=1,2,3,4,5)$. Edges correspond to triangles carrying spins $j_{ab}$. Circles as endpoints of edges carry boundary states $\xi_{ab}$ and $\xi_{ba}$. Arrows represent orientations $a<b$.}
\label{4simplex}
\end{figure}

Here we focus on the Lorentzian 4-dimensional spinfoam 4-simplex amplitude, illustrated by FIG.\ref{4simplex}, where each black box is dual to a boundary tetrahedron and each edge is dual to a triangle. Boundary tetrahedra are labeled by indexes $a,b=1,...,5$ and carry group variables $g_a\in\Slc$. The triangle dual to the edge is shared by the $a$-th and $b$-th tetrahedra and carries an SU(2) spin $j_{ab}$. We firstly impose the boundary state made by a tensor product of five coherent intertwiners, one for each tetrahedron,
\be
|i_a (\vec{j},\vec{\xi})\rangle=\int_{\Su}\rmd h_a\bigotimes_{b\neq a} h_a|j_{ab},\xi_{ab}\rangle,\quad a=1,\cdots,5,
\ee
where $|j_{ab},\xi_{ab}\rangle$ is the SU(2) coherent state. The EPRL 4-simplex amplitude with the boundary state has the integral expression \cite{Conrady:2008ea,Conrady:2008mk,Barrett:2009gg,Barrett:2009mw,Han:2011re,Han:2011rf} which is  particularly useful for studying the large-$j$ asymptotic behavior. 
\begin{equation}
A_v (j_{ab},i_a)= \int \prod_{a=2}^{5} dg_a \int_{(\mathbb{CP}^1)^{10}} e^S \prod_{a<b}\frac{d_{j_{ab}}}{\pi}\Omega_{ab}, \label{ampaction}
\end{equation}
with $d_{j_{ab}}=2j_{ab}+1$. $g_a\in\mathrm{SL}(2, \mathbb{C})$ associates to each tetrahedron. The first tetrahedron is gauge fixed, $g_1=\ident$. $dg_a$ is the Haar measure on $\Slc$. $\Omega_{ab}$ is the measure on $\mathbb{CP}^1$:
\begin{equation}
	\Omega_{ab}=\frac{\Omega}{\left\langle Z_{ab}, Z_{ab}\right\rangle\left\langle Z_{ba}, Z_{ba}\right\rangle},\label{Omegaab}
\end{equation}
where $\left| Z_{ab}\right\rangle = g_{a}^{\dagger} \left|z_{ab}\right\rangle$, $\left| Z_{ba}\right\rangle = g_{b}^{\dagger} \left|z_{ab}\right\rangle$ and $\left|z_{ab}\right\rangle$ is a 2-component spinor for each triangle $ab$. The Hermitian inner product is $\langle z, w\rangle=\overline{z}_{0} w_{0}+\overline{z}_{1} w_{1}$. Here $\Omega=\frac{i}{2}\left(z_0dz_1-z_1dz_0\right)\wedge\left(\bar{z}_0d\bar{z}_1-\bar{z}_1d\bar{z}_0\right)$ is a homogeneous measure on $\mathbb{C}^2$, and we choose the section of $\mathbb{CP}^1$: $\left(z_0,z_1\right)\rightarrow \left(-\sin\frac{\Theta}{2}\e^{-i\Phi},\cos\frac{\Theta}{2}\right)$, for which $\Omega$ reduces to $\Omega=\frac{\sin\Theta}{4}d\Phi d\Theta$.
 
The integrand in Eq.(\ref{ampaction}) is written as an exponential $e^S$ with the action
\begin{equation}
S=\sum_{a<b} 2j_{ab} \ln \frac{\left\langle Z_{ab},\xi_{ab}\right\rangle\left\langle J\xi_{ba},Z_{ba}\right\rangle}{\left\langle Z_{ab}, Z_{ab}\right\rangle^{\frac{1}{2}}\left\langle Z_{ba}, Z_{ba}\right\rangle^{\frac{1}{2}}}+i \gamma j_{ab} \ln \frac{\left\langle Z_{ba}, Z_{ba}\right\rangle}{\left\langle Z_{ab}, Z_{ab}\right\rangle}, \label{action0}
\end{equation} 
where $\gamma$ is the Barbero-Immizi parameter and $J$ is the anti-linear map 
$$
J\left(\begin{matrix}
z_0\\z_1
\end{matrix}\right)=\left(\begin{matrix}
-\bar{z}_1\\\bar{z}_0
\end{matrix}\right).
$$
The coherent state is labeled by the spin $j_{ab}$ and a normalized 2-component spinor $|\xi_{ab}\rangle$ which is determined by $\hat{n}_{ab}=\langle \xi_{ab},\vec{\sig}\xi_{ab}\rangle$ ($\vec{\sig}$ are Pauli matrices).

To study the large-$j$ behavior of the amplitude, we scale spins $j_{ab}\rightarrow\lambda j_{ab}$ by a large parameter $\lambda$. As a consequence of the scaling of spins, the action $S\mapsto\lambda S$. This scaling motivates us to study the asymptotical behavior of $A_v$ in the large-$j$ regime with the generalized stationary phase approximation analysis guided by H\"ormander's theorem 7.7.5 \cite{Hormander}.

\begin{Theorem}\label{hormanderThm}

Let $K$ be a compact subset in $\mathbb{R}^n$, X an open neighborhood of K, and k a positive integer.  If (1) the complex functions $u\in C^{2k}_{0}(K)$, $f\in C^{3k+1}(X)$ and Im $f\geq 0$ in X; (2) there is a unique point $x_0\in K$ satisfying $\mathrm{Im}(S(x_0))=0$, $f'(x_0)=0$, and $\det (f''(x_0))\neq 0$ ($f''$ denotes the Hessian matrix), $f'\neq 0$ in $K \backslash\left\{x_{0}\right\}$ then we have the following estimation:\\
\begin{equation}
\left|\int_{K} u(x) e^{i\lambda f(x)} dx-e^{i\lambda f(x_0)}\left[\mathrm{det}\left(\dfrac{\lambda f''(x_0)}{2\pi i}\right)\right]^{-\frac{1}{2}}\sum_{s=0}^{k-1}\left(\dfrac{1}{\lambda}\right)^s L_s u(x_0)\right|\leq C\left(\dfrac{1}{\lambda}\right)^{k}
\sum_{|\alpha|\leq 2k}\mathrm{sup}\left|D^{\alpha}u\right|.\label{therom}
\end{equation}
Here the constant C is bounded when f stays in a bounded set in $C^{3k+1}(X)$. We have used the standard multi-index notation $\alpha=\langle \alpha_1,...,\alpha_n\rangle$ and
\begin{equation}
D^{\alpha}=(-i)^\alpha\frac{\partial^{|\alpha|}}{\partial x_1^{\alpha_1}...\partial x_n^{\alpha_n}}, \quad \text{where}\quad |\alpha|=\sum_{i=1}^{n}\alpha_i
\end{equation}
$L_s u(x_0)$ denotes the following operation on u:
\begin{equation}
L_s u(x_0)= i^{-s} \sum_{l-m=s}\sum_{2l\geq 3m}\frac{(-1)^l 2^{-l}}{l!m!}\left[\sum_{a,b=1}^{n}H_{ab}^{-1}(x_0)\dfrac{\partial^2}{\partial x_a\partial x_b}\right]^l \left(g_{x_0}^m u\right)\left(x_0\right),\label{theorem}
\end{equation}
where $H(x)=f''(x)$ denotes the Hessian matrix and the function $g_{x_0}(x)$ is given by
\begin{equation*}
g_{x_0}(x)=f(x)-f(x_0)-\frac{1}{2}H^{ab}\left(x_0\right)\left(x-x_0\right)_a\left(x-x_0\right)_b
\end{equation*}
such that $g_{x_0}\left(x_0\right)=g'_{x_0}\left(x_0\right)=g''_{x_0}\left(x_0\right)=0$. For each s, $L_s$ is a differential operator of order 2s acting on $u\left(x\right)$.

\end{Theorem}

Employing this Theorem, we can compute the 4-simplex amplitude in Eq.(\ref{ampaction}) as an $1/\lambda$ asymptotic series at critical points. As a consequence, the asymptotics of 4-simplex amplitude as $\lambda\rightarrow\infty$ is dominated by contributions of critical points which are the solutions of the critical point equations, 
\begin{equation}
\mathrm{Re}(S)=0,\ \partial_{z_{ab}} S=0,\ \mbox{and}\ \partial_{g_a} S=0,
\label{eq:criticalpt}
\end{equation}
where $S[g,z]$ is given by Eq.(\ref{action0}). Results from literatures  \cite{Barrett:2009gg,Barrett:2009mw,Freidel:2007py,Han:2011re,Han:2011rf} show that for boundary states whose data $j_{ab},\xi_{ab}$ correspond to the geometrical boundary of a non-degenerate 4-simplex (and satisfy the orientation matching condition), $S$ has two critical points having the geometrical interpretation as the non-degenerate geometrical 4-simplex with opposite orientations. $S$ evaluated at critical points gives the Regge action of the 4-simplex with opposite signs. In the next section, we review the boundary data and the construction of critical points for the EPRL amplitude with the coherent intertwiner boundary state.

\section{Boundary data and critical point} \label{bd}
\subsection{Boundary data}

The boundary state $\ket{\psi}=\otimes_{a=1}^5|i_a\rangle$ for demonstrating our algorithm is the same as \cite{Dona:2019dkf}. $\ket{\psi}$ is labeled by ten spin variables $\l j_{ab}$ and twenty $\xi_{ab}$ which relate to face 3-d normals $\vec{n}_{ab}$. As an example, we set the area of six faces of the geometrical 4-simplex to be 2 and other areas to be 5. We focus on this example in this section, a few other examples is given in Appendix \ref{boundary geometry}. Although we use dimensionless numbers to describe the areas, physical areas are obtained by attaching proper units to those numbers. In our calculation, those areas are $j_{ab}$ (spins are $\l j_{ab}$). Furthermore, the face normals, denoted as $\vec{n}_{ab}$, are gained by the 4-simplex geometry. For convenience, we denote the five vertices of the 4-simplex as $P_a$ and five tetrahedra as $T_a$, where $a\in\{1,2,3,4,5\}$. We firstly write down the coordinates of the vertices $P_a$ in the Minkowski spacetime. Our starting point is the tetrahedra $T_1$, which is an equilateral tetrahedron with all areas equaling to 5. We endow the vertices of $T_1$ with coordinates $P_1=(0, 0, 0, 0)$, $P_2=(0, 0, 0, -2 \sqrt{5}/3^{1/4})$, $P_3=(0, 0, -3^{1/4} \sqrt{5},-3^{1/4} \sqrt{5})$ and $P_4=(0, -2 \sqrt{10}/3^{3/4}, -\sqrt{5}/3^{3/4}, -\sqrt{5}/3^{1/4})$ respectively. It means that we locally set up a frame $(t,x,y,z)$ so that $T_1$ is embedded in the subspace expanded by $x,y,z$ axis. The 4-simplex can be well located in our frame if one can find a coordinate of the vertex $P_5=(t_1,x_1,y_1,z_1)$ such that the 4-d distances between $P_5$ and $P_a\  (a\neq 5)$ are the same and areas of the triangles connecting $P_5$ to other $P_a$ are all 2. By solving the system of equations, one can find $P_5$ is $(-3^{-1/4} 10^{-1/2}, -\sqrt{5/2}/3^{3/4}, -\sqrt{5}/3^
{3/4}, -\sqrt{5}/3^{1/4})$. Then, from the coordinates of $P_a$, we calculate the 4-d normals $N_a$ of each tetrahedra $T_a$ respectively. From the vertices, we compute the edge vectors $l_{ae}^I$ of the tetrahedron $a$ at edge $e$, with $I=0,1,2,3$ a Cartesian coordinate index. Then one can determine the 4-d normals, $N_{a}$, from the triple product of wedges with a common vertex determined by three edges labeled by $e=1,2,3$ respectively
$$
N_{aI}=\frac{\epsilon_{IJKL}l_{a1}^J l_{a2}^Kl_{a3}^L}{\left\|\epsilon_{IJKL}l_{a1}^Jl_{a2}^Kl_{a3}^L\right\|},
$$
where the norms and scalar products are given by the Minkowski metric $\eta=\text{diag}(-,+,+,+)$, and the epsilon symbol is of the convention $\epsilon_{0123}=1$. Hence, 4-d normal vectors are given by:
\begin{equation}\label{eq:4Normals}
\begin{split}
N_1=&(-1,0,0,0),\ 
N_2=(\f{5}{\sqrt{22}},\sqrt{\f{3}{22}},0,0),\ 
N_3=(\f{5}{\sqrt{22}},-\f{1}{\sqrt{66}},\f{2}{\sqrt{33}},0),\\ 
N_4&=(\f{5}{\sqrt{22}},-\f{1}{\sqrt{66}},-\f{1}{\sqrt{33}},\f{1}{\sqrt{11}}),\ 
N_5=(\f{5}{\sqrt{22}},-\f{1}{\sqrt{66}},-\f{1}{\sqrt{33}},-\f{1}{\sqrt{11}}).\\ 
\end{split} 
\end{equation}
The next step is to find the transformation which takes all 4-d normal vectors to the time gauge $T=(-1,0,0,0)$ \cite{Dona:2019dkf}:
$$
\Lambda_{a  J}^{I}=\eta^I_J+\frac{1}{1-N_a\cdot T}\left(N_a^IN_{aJ}+T^IT_{J}+N_a^{I}T_J-\left(1-2N_a\cdot T\right)T^IN_{aJ}\right),
$$
$$
\Lambda_{a  J}^{I}N^{J}_a=T^{I}\quad \det\Lambda_{a  J}^{I}=1,\quad a\neq 1,\quad I,J=0,1,2,3.
$$ 
Then the 3-d face normals are 
\begin{equation}
	n_{ab}^I:=-\Lambda_{a  J}^{I}\frac{N_b^{J}+N_a^{J}\left(N_a\cdot N_b\right)}{\sqrt{\left(N_a\cdot N_b\right)^2-1}}.\label{3dnormal}
\end{equation}
The gauge-fixed tetrahedron, $a=1$, has $\Lambda_{1}=\eta$ and $N_1=T$. The 3-d normals resulting from Eq.(\ref{3dnormal}) are showing in Table \ref{tab:3dNormal}. 

\begin{table}[h]
	\centering\caption{Each cell of the table is the 3-d normal vector for the face shared by the line number tetrahedra and the column number tetrahedra.}\label{tab:3dNormal}
	\small
	\setlength{\tabcolsep}{0.8mm}
	\begin{tabular}{|c|c|c|c|c|c|}
		\hline
		\diagbox{\small{a}}{normal $\vec{n}_{ab}$ }{\small{b}}&1&2&3&4&5\\
		\hline
		1&\diagbox{}{}&(1,0,0)&(-0.33,0.94,0)&(-0.33,-0.47,0.82)&(-0.33,-0.47,-0.82)\\
		\hline
		2&(-1,0,0)&\diagbox{}{}&(0.83,0.55,0)&(0.83,-0.28,0.48)&(0.83,-0.28,-0.48)\\
		\hline
		3&(0.33,-0.94,0)&(0.24,0.97,0)&\diagbox{}{}&(-0.54,0.69,0.48)&(-0.54,0.69,-0.48)\\
		\hline
		4&(0.33,0.47,-0.82)&(0.24,-0.48,0.84)&(-0.54,0.068,0.84)&\diagbox{}{}&(-0.54,-0.76,0.36)\\
		\hline
		5&(0.33,0.47,0.82)&(0.24,-0.48,-0.84)&(-0.54,0.068,-0.84)&(-0.54,-0.76,-0.36)&\diagbox{}{}\\
		\hline
	\end{tabular}
\end{table}

\noindent $\vec{n}_{ab}$ can be converted to the spinor $\ket{\xi_{ab}}$ (by fixing the phase convention):
\begin{equation}
\vec{n}_{ab}= (\sin\Theta\cos\Phi,\sin\Theta\sin\Phi,\cos\Theta) \rightarrow \ket{\xi_{ab}}= \left(-\sin\frac{\Theta}{2}\e^{-i\Phi},\cos\frac{\Theta}{2}\right).
\end{equation}
The boundary state $\ket{\xi_{ab}}$ is showing in Table
\ref{tab:xi}. Once boundary data $j_{ab},\xi_{ab}$ are fixed, critical points ($g_{a}^0$, $z_{ab}^0$) are obtained by solving critical point equations (\ref{eq:criticalpt}).

\begin{table}[h]
    \centering\caption{Each cell of the table is boundary state $\ket{\xi_{ab}}$ for the face shared by the line number tetrahedra and the column number tetrahedra.}\label{tab:xi}
	\footnotesize
	\setlength{\tabcolsep}{0.8mm}
	\begin{tabular}{|c|c|c|c|c|c|}
		\hline
		\diagbox{\small{a}}{$\ket{\xi_{ab}}$ }{\small{b}}&1&2&3&4&5\\
		\hline
		1&\diagbox{}{}&(0.71,0.71)&(0.71,-0.24+0.67 i)&(0.95,-0.17-0.25 i)&(0.30,-0.55-0.78 i)\\
		\hline
		2&(0.71,-0.71)&\diagbox{}{}&(0.71,0.59+0.39 i)&(0.86, 0.48 - 0.16 i)&(0.51, 0.82 - 0.27 i)\\
		\hline
		3&(0.71, 0.24 - 0.67 i)&(0.71, 0.17 + 0.69 i)&\diagbox{}{}&(0.86, -0.31 + 0.40 i)&(0.51, -0.53 + 0.68 i)\\
		\hline
		4&(0.30, 0.55 + 0.78 i)&(0.96, 0.13 - 0.25 i)&(0.96, -0.28 + 0.035 i)&\diagbox{}{}&(0.83, -0.33 - 0.46 i)\\
		\hline
		5&(0.95, 0.17 + 0.25 i)&(0.28, 0.43 - 0.86 i)&(0.28, -0.95+ 0.12 i)&(0.57, -0.48-0.67 i)&\diagbox{}{}\\
		\hline
	\end{tabular}
\end{table}

\subsection{Critical points}
Critical points of the integral \eqref{ampaction} are denoted by ($g_{a}^0$, $z_{ab}^0$). From the critical point equations (\ref{eq:criticalpt}), $\mathrm{Re}(S)=0$ leads to the equations \cite{Barrett:2009gg}
\be
\ket{\xi_{ab}}=\frac{e^{i\psi _{ab}}}{\left\Vert Z_{ab}\right\Vert }g^{\dagger}_{a}\ket{z_{ab}},\quad
\text{and}\quad \ket{J\xi_{ba}}=\frac{e^{i\psi _{ba}}}{\left\Vert Z_{ba}\right\Vert }g^{\dagger}_{a}\ket{z_{ab}},  \label{eq:ReS}
\ee
where $\left\Vert Z_{ab}\right\Vert \equiv \left\vert \left\langle
Z_{ab},Z_{ab}\right\rangle \right\vert ^{1/2}$, $\psi_{ab}$ and $\psi_{ba}$ are phases. The two equations  above can be combined to 
\be
(g^{\dagger}_{a})^{-1}\ket{\xi_{ab}}=\frac{\left\Vert Z_{ba}\right\Vert}{\left\Vert Z_{ab}\right\Vert}e^{i(\psi_{ab}-\psi_{ba})}(g^{\dagger}_{b})^{-1}\ket{J\xi_{ba}}. \label{eq:ReS1}
\ee
The variation of the action with respect to a spinor $\partial_{z_{ab}} S=0$ leads to the equation 
\be
g_{a}\ket{\xi_{ab}}=\frac{\left\Vert Z_{ab}\right\Vert}{\left\Vert Z_{ba}\right\Vert}e^{i(\psi_{ab}-\psi_{ba})}g_{b}\ket{J\xi_{ba}}.\label{eq:deltaZ}
\ee
The variation with respect to $g_a$ gives the closure condition
\be
\sum_b j_{ab}\vec{n}_{ab}=0.
\ee
Solutions of above equations have been studied extensively in the literature. Given the boundary data $j_{ab},\xi_{ab}$ which satisfies the orientation matching condition, then above equations have two solutions corresponding to the 4-simplex geometry with opposite orientations \cite{Barrett:2009mw} . These solutions are denoted by $(g^{0(\pm)},z^{0(\pm)})$. $g^{0(\pm)}$ relates to the Lorentz transformation acting on each tetrahedron $T_a$ and gluing them together to form one 4-simplex. They can be expressed explicitly \cite{Dona:2019dkf} by $g_1^0=\ident$, 
\be
g_a^{0(\pm)}=\exp\left((\pm\theta_{1a}^L+i\pi)\vec{n}_{1a}\cdot\frac{\vec{\sigma}}{2}\right), \quad a\neq 1, \label{criti:ga}
\ee
where $\vec{\sigma}$ are Pauli matrices. $\theta_{1a}^L$ is the 4-d dihedral angle which is the boost angle between two 4-d normals of tetrahedra, defined by:
\be
N_1\cdot N_a=\cosh\theta_{1a}^{L}, \quad a\neq 1,
\ee
where $N_1$ and $N_a$ are given by (\ref{eq:4Normals}). From Eq. (\ref{criti:ga}), we can see that $g_a^{0(\pm)}$ are combinations of boosts given by $\pm\theta_{1a}^{L}$ and an additional rotation $\pi$ in the same direction. Resulting from this rotation, 3-d normals in the first tetrahedron are opposite to the corresponding ones in the adjacent tetrahedra. The numerical results for critical point $g_a^{0(\pm)}$ are shown in Table \ref{tab:ga}. 

\begin{table}[h]
	\centering\caption{Each cell of the table is the critical point of a-th tetrahedron group element $g_a^{0(\pm)}$.}\label{tab:ga}
	\scriptsize
	\setlength{\tabcolsep}{0.5mm}
	\begin{tabular}{|c|c|c|c|c|c|}
		\hline
		\small{a}&1&2&3&4&5\\
		\hline
		$g_a^{0(+)}$ &$\left(\begin{matrix}
		1&0\\	0&1
		\end{matrix}\right)$&$\left(\begin{matrix}
		0.18 i&1.01 i\\	1.01 i&0.18 i
		\end{matrix}\right)$&$\left(\begin{matrix}
		0.18 i&0.96 - 0.34 i\\	-0.96- 0.34 i&0.18 i
		\end{matrix}\right)$&$\left(\begin{matrix}
		1.01 i&-0.48 - 0.34 i\\	0.48 - 0.34 i&-0.65 i
		\end{matrix}\right)$&$\left(\begin{matrix}
		-0.65 i&-0.48 - 0.34 i\\	0.48 - 0.34 i&1.01 i
		\end{matrix}\right)$\\
		\hline
		$g_a^{0(-)}$ &$\left(\begin{matrix}
		1&0\\	0&1
		\end{matrix}\right)$&$\left(\begin{matrix}
		-0.18 i&-1.01 i\\	1.01 i&-0.18 i
		\end{matrix}\right)$&$\left(\begin{matrix}
		-0.18 i&0.96 - 0.34 i\\	-0.96- 0.34 i&-0.18 i
		\end{matrix}\right)$&$\left(\begin{matrix}
		0.65 i&-0.48 - 0.34 i\\	0.48 - 0.34 i&-1.01 i
		\end{matrix}\right)$&$\left(\begin{matrix}
		-1.01 i&-0.48 - 0.34 i\\	0.48 - 0.34 i&0.65 i
		\end{matrix}\right)$\\
		\hline
	\end{tabular}
\end{table} 

$z_{ab}$ can be fixed by $g_a^{0(\pm)}$ and $\ket{\xi_{ab}}$, up to a complex scaling, by the first equation in Eqs.(\ref{eq:ReS})
\be
\ket{z_{ab}^{0(\pm)}}\propto_\C (g_{a}^{0(\pm)\dagger})^{-1} \left|\xi_{ab}\right\rangle.
\ee
Scaling of $z_{ab}$ is a gauge transformation of $S$. We fix the scaling by normalizing $\ket{z_{ab}^{0(\pm)}}$ and making the following parametrization 
\begin{equation}
	\ket{z^{0(\pm)}_{ab}}=\left(\begin{matrix}
	-\sin\frac{\theta^{0(\pm)}_{ab}}{2}\, e^{-i\phi_{ab}^{0(\pm)}}\\ \cos\frac{\theta_{ab}^{0(\pm)}}{2}
	\end{matrix}\right),\label{z0}
\end{equation} 
here, $\theta_{ab}^{0(\pm)}$ and $\phi_{ab}^{0(\pm)}$ are real. 

The numerical results for $(\theta_{ab}^{0(\pm)}, \phi_{ab}^{0(\pm)})$ are shown in Table \ref{tab:pos}. All critical point data of $z_{ab}^{0(\pm)}$ and $g_{a}^{0(\pm)}$ can be found in Mathematica notebooks \cite{qudx.org}. 

\begin{table}[h]
    \centering\caption{Each cell of the table is the critical point parameterized by $(\theta_{ab}^{0(\pm)}, \phi_{ab}^{0(\pm)})$ for the face shared by the line number tetrahedron a and the column number tetrahedron b, $a<b$. Two tables list the results for two distinct critical points.}\label{tab:pos}
	\footnotesize
	\setlength{\tabcolsep}{0.8mm}
	\begin{tabular}{|c|c|c|c|c|}
		\hline
		\diagbox{\small{a}}{$(\theta_{ab}^{0(+)}, \phi_{ab}^{0(+)})$ }{\small{b}}&2&3&4&5\\
		\hline
		1&(-1.57,0)&(-1.57,1.91)&(-2.53,-2.19)&(-0.62,-2.19)\\
		\hline
		2&\diagbox{}{}&(-1.57,-0.82)&(-0.89, 0.49)&(-2.25, 0.49)\\
		\hline
		3&\diagbox{}{}&\diagbox{}{}&(-0.89, 1.42)&(-2.25, 1.42)\\
		\hline
		4&\diagbox{}{}&\diagbox{}{}&\diagbox{}{}&(-2.94, 0.96)\\
		\hline
	\end{tabular}
	\begin{tabular}{|c|c|c|c|c|}
		\hline
		\diagbox{\small{a}}{$(\theta_{ab}^{0(-)}, \phi_{ab}^{0(-)})$ }{\small{b}}&2&3&4&5\\
		\hline
		1&(-1.57,0)&(-1.57,1.91)&(-2.53,-2.19)&(-0.62,-2.19)\\
		\hline
		2&\diagbox{}{}&(-1.57,-0.41)&(-1,21, 0.22)&(-1.93, 0.22)\\
		\hline
		3&\diagbox{}{}&\diagbox{}{}&(-1.21, 1.69)&(-1.93, 1.69)\\
		\hline
		4&\diagbox{}{}&\diagbox{}{}&\diagbox{}{}&(-2.94, -2.19)\\
		\hline
	\end{tabular}
\end{table}
In this work, we choose a specific boundary geometry $j_{ab}=(5,2)\lambda$ to be consistent with the results of \cite{Dona:2019dkf}. To improve the strength of our claims, we also perform the analysis with other boundary geometries $j_{ab}=(8,3)\l$ and $j_{ab}=(11,4)\l$ in the same way. The boundary data and critical points for boundary geometries  $j_{ab}=(8,3)\l$ and $j_{ab}=(11,4)\l$ can be found in Appendix \ref{boundary geometry}. In the following context, if we do not specify the boundary geometry particularly, the boundary geometry we are referring to is $j_{ab}=(5,2)\l$.     

\section{Next-to-leading order correction in large-$j$ 4-simplex amplitude with coherent-intertwiner boundary state}\label{intertwiner}

\subsection{Explicit expression of 4-simplex amplitude}

Given one of the critical points, we make the following parametrization of $g_a$ and $z_{ab}$ of the neighborhood of the critical point in the integration domain of Eq.(\ref{ampaction}):
The group variable $g_{a} \in \mathrm{SL}(2, \mathbb{C})$ is parametrized by
\begin{equation}
g_{a}\left(x_{a1},y_{a1},x_{a2},y_{a2},x_{a3},y_{a3}\right)=g_{a}^{0(\pm)}\left(\begin{matrix}1+\frac{x_{a1}+i y_{a1}}{\sqrt{2}}&\frac{x_{a2}+i y_{a2}}{\sqrt{2}}\\\frac{x_{a3}+i y_{a3}}{\sqrt{2}}&\frac{1+\frac{x_{a2}+i y_{a2}}{\sqrt{2}}\frac{x_{a3}+i y_{a3}}{\sqrt{2}}}{1+\frac{x_{a1}+i y_{a1}}{\sqrt{2}}}\end{matrix}\right)
\label{ge},
\end{equation}
where $x_{ai}$ and $y_{ai} (a\neq 1, i=1,2,3)$ are real. The first tetrahedron is gauge-fixed, $g_1=\ident$. There are 24 real variables $x_{ai}$ and $y_{ai} (a=2,3,4,5)$ because of group variables $g_a$. The spinor $z_{ab}\in \mathbb{CP}^{1}$ is parametrized by
\begin{equation}
z_{ab}\left(\Theta_{ab},\Phi_{ab}\right)=\begin{pmatrix} 
-\sin\left(\frac{\theta^{0(\pm)}_{ab}+\Theta_{ab}}{2}\right)\e^{-i(\phi^{0(\pm)}_{ab}+\Phi_{ab})} \\
\cos\left(\frac{\theta_{ab}^{0(\pm)}+\Theta_{ab}}{2}\right)
\end{pmatrix}.\label{zef}
\end{equation}
Each triangle $ab$ has two real variables $\Theta_{ab}$ and $\Phi_{ab}$, so we need in total 20 real variables to describe $z_{ab}$. $\ket{Z_{ab}}=g_{a}^{\dagger} z_{ab}$ follows from (\ref{ge}) and (\ref{zef}). The arguments of the action in Eq.(\ref{action0}) are now $x_{ai}$, $y_{ai}$, $\Theta_{ab}$ and $\Phi_{ab}$.

For the group integral, the $\mathrm{SL}(2, \mathbb{C})$ Haar measure $dg$ can be written explicitly by
\begin{equation}
\prod_{a=2}^{5}dg_a=\prod_{a=2}^{5} \frac{1}{2^7\pi^4}\frac{dx_{a1} dy_{a1} dx_{a2} dy_{a2} dx_{a3} dy_{a3}}{\left|1+\frac{x_{a1}+i y_{a1}}{\sqrt{2}}\right|^2}.\label{haar}
\end{equation}
The details of this derivation are in appendix \ref{Haar measure}. We define the function $u\left(x_{ai},y_{ai},\Theta_{ab},\Phi_{ab}\right)$ by
\begin{equation}
u\left(x_{ai},y_{ai},\Theta_{ab},\Phi_{ab}\right) \prod_{a<b}\prod_{i=1,2,3}dx_{ai}dy_{ai}d\Phi_{ab}d\Theta_{ab} = \prod_{a<b} \Omega_{ab} dg_{a},
\end{equation}
where $\Omega_{ab}$ is the measure on $\mathbb{CP}^1$ in (\ref{Omegaab}):
\begin{equation}
	\Omega_{ab}=\frac{\sin\left(\theta^{0(\pm)}_{ab}+\Theta_{ab}\right)}{4\langle Z_{ab},Z_{ab}\rangle\langle Z_{ba},Z_{ba}\rangle}d\Theta_{ab} d\Phi_{ab}.\label{Omega}
\end{equation}
There are 44 arguments in $u\left(x_{ai},y_{ai},\Theta_{ab},\Phi_{ab}\right)$. The amplitude (\ref{ampaction}) gives
\begin{equation}
\begin{aligned}
 \prod_{a<b} \frac{d_{\lambda j_{ab}}}{\pi} \int \prod_{i=1}^{44}dx_i u\left(\vec{\bold{x}}\right) e^{\l S\left(\vec{\bold{x}}\right)}, 
\label{explicitamp}
\end{aligned}
\end{equation}
where $\vec{\bold{x}}=(x_1,x_2,\cdots,x_{44})\equiv \left(x_{ai},y_{ai},\Theta_{ab},\Phi_{ab}\right)$ contains 44 components. Eq.(\ref{explicitamp}) expresses the EPRL 4-simplex amplitude in the form as Eq.(\ref{theorem}). Besides, the critical point for $S(\vec{\bold{x}})$ is at $\vec{\bold{x}}_0=(0,0,..,0)$. Next, we will apply Theorem \ref{hormanderThm} to expand the integral (\ref{explicitamp}) and numerically compute the leading-order terms and next-to-leading order corrections.

We emphasize that the parametrization $\vec{\bf x}$ is within the neighborhood of \emph{one} critical point in the integration domain, and (\ref{explicitamp}) is $A_v$ restricted in the neighborhood. $A_v$ has 2 critical points which leads to 2 different notions of $\vec{\bf x}$. We don't put the label $(\pm)$ to $\vec{\bf x}$ in order to make notations less cumbersome.

\subsection{Asymptotic expansion and next-to-leading order correction} \label{Asymptotic Expansion with the coherent-intertwiner boundary state}

Following the convention in Theorem \ref{hormanderThm}, we rewrite the exponent in the integrand  
\begin{equation*}
\int \prod_{i=1}^{44}dx_i u\left(\vec{\bold{x}}\right)  e^{\lambda S\left(\vec{\bold{x}}\right)}=\int \prod_{i=1}^{44}dx_i u\left(\vec{\bold{x}}\right)  e^{i\lambda \tilde{S}\left(\vec{\bold{x}}\right)}.
\end{equation*}
Here, $\tilde{S}\left(\vec{\bold{x}}\right)=-iS\left(\vec{\bold{x}}\right)$, Hessian matrix $H_{ij}(\vec{\bold{x}})=\partial_i\partial_j\tilde{S}(\vec{\bold{x}})$. The leading-order terms and the next-to-leading-order corrections in Eq.(\ref{theorem}) correspond to $s=0$ and $s=1$. In (\ref{theorem}), the expression of $L_s u\left(x_0\right)$ sums a finite number of terms for each $s$. 

Our scheme of computation is as follows: at $s=0$, the corresponding term for $L_{s=0}  u\left(\boldsymbol{\vec{\bold{x}}}\right)$ is
\begin{equation}
I_0=u(0).	
\end{equation}

At $s=1$, the possible $(m,l)$ are $(0,1),(1,2),(2,3)$ to satisfy $2l\geq 3m$. The corresponding terms are of the types:
\begin{itemize}
	\item[1)]
$\left(m,l\right)=\left(0,1\right)$:
\begin{equation}
I_1=-\dfrac{1}{2i}\left[\sum_{i,j=1}^{44} H_{ij}^{-1}\left(0\right)\dfrac{\partial^2 u\left(0\right)}{\partial x_i\partial x_j}\right] \label{I1}.
\end{equation} 
where we have expressed $\vec{\bf x}=(x_1,x_2,\cdots,x_{44})$. We compute the second-order derivative of the function $u\left(\vec{\bold{x}}\right)$ with respect to $x_i$ and $x_j$ and evaluate the result at $\vec{\bf x}=0$. We save the resulting 44$\times$44 matrix $\dfrac{\partial^2 u\left(0\right)}{\partial x_i\partial x_j}$ and contract it with the Hessian matrix.\\
	\item[2)]
$\left(m,l\right)=\left(1,2\right)$: We define 
\be
g_{x_{0}}(\vec{\bf x})=\tilde{S}(\vec{\bf x})-\tilde{S}\left(0\right)-\frac{1}{2}\sum_{i,j=1}^{44} H_{ij}\left(0\right)x_ix_j,
\ee
\begin{equation}
\begin{aligned}
I_2=&\dfrac{1}{8i}\left[\sum_{i,j=1}^{44} H_{ij}^{-1}\left(0\right)\dfrac{\partial^2 }{\partial x_i\partial x_j}\right]\left[\sum_{k,l=1}^{44} H_{kl}^{-1}\left(0\right)\dfrac{\partial^2 }{\partial x_k\partial x_l}\right]\left(g_{x_0}u\right)(0)\\=&\dfrac{1}{8i}\left[\sum_{i,j,k,l=1}^{44}H_{ij}^{-1}H_{kl}^{-1}\dfrac{\partial^4}{\partial x_i\partial x_j\partial x_k\partial x_l}\right]\left(g_{x_0}u\right)(0)\\=&\dfrac{1}{8i}\sum_{i,j,k,l=1}^{44}H_{ij}^{-1}H_{kl}^{-1}\left[\dfrac{\partial^3 g_{x_0}(0)}{\partial x_i\partial x_j\partial x_k}\dfrac{\partial u(0)}{\partial x_l}+\dfrac{\partial^3 g_{x_0}(0)}{\partial x_i\partial x_j\partial x_l}\dfrac{\partial u(0)}{\partial x_k}+\dfrac{\partial^3 g_{x_0}(0)}{\partial x_j\partial x_k\partial x_l}\dfrac{\partial u(0)}{\partial x_i}\right.\\\quad&\quad\quad\quad+\dfrac{\partial^3 g_{x_0}(0)}{\partial x_i\partial x_k\partial x_l}\dfrac{\partial u(0)}{\partial x_j}+\left.\dfrac{\partial^4 g_{x_0}(0)}{\partial x_i\partial x_j\partial x_k\partial x_l}u(0)\right].
\end{aligned}\label{I2}
\end{equation}
To compute less expensively, we use following techniques in our code \cite{qudx.org}. First, 
for $\partial ^3 g_{x_0}(0)$ and $\partial ^4 g_{x_0}(0)$, there are $44^{3}$ possible third order partial derivatives and $44^{4}$ fourth order partial derivatives. However, for function $g_{x_0}(\vec{\bf x})$, the mixed partial derivatives are equal, which means that the order in which we differentiate won't matter. To save space, we save $\partial ^3 g_{x_0}(0)$ as a 3-d upper ``triangle" array $G^{(3)}$ and $\partial ^4 g_{x_0}(0)$ as a 4-d upper ``triangle" array $G^{(4)}$. Hence, the size of the array is deduced from $44^d$ to $\left(\begin{matrix}
44+d-1\\d
\end{matrix}\right)$.  Secondly, because of the symmetric property of the Hessian matrix, we can simplify the above summation from $\sum_{i,j,k,l=1}^{44}F_{ijkl}$ to $\sum_{i=1}^{44}\sum_{j=1}^{i}\sum_{k=1}^{j}\sum_{l=1}^{k}\tilde{F}_{ijkl}$, here we use $F_{ijkl}$ to denote the factors' product in Eq.(\ref{I2}). To be clear, we use an example $i=9,j=8,k=8,l=1$, i.e., $\left(9,8,8,1\right)$, to explain the technique. We sum all possible $H^{-1}_{ij}H^{-1}_{kl}$ first, it's 
$4H^{-1}_{19}H^{-1}_{88}+8H^{-1}_{81}H^{-1}_{89}$. Here, we use the symmetric property of $H^{-1}$ and the counting principle to get each terms and the corresponding coefficient. In this case,
\begin{equation}
\begin{aligned}
\tilde{F}_{9881}=&\left(4H^{-1}_{19}H^{-1}_{88}+8H^{-1}_{81}H^{-1}_{89}\right)\\
&\times\left(G^{(3)}_{988}\frac{\partial u(0)}{\partial x_1}+2\times G^{(3)}_{981}\frac{\partial u(0)}{\partial x_8}+G^{(3)}_{881}\frac{\partial u(0)}{\partial x_9}+G^{(4)}_{9881}u(0)\right).
\end{aligned}
\end{equation}
One can find the details in our Mathematica notebooks \cite{qudx.org}. 
\item[3)] $\left(m,l\right)=\left(2,3\right)$: 
\begin{equation}
\begin{aligned}
I_3=-&\dfrac{1}{96i}\left[\sum_{i,j,k,l,m,n=1}^{44}H_{ij}^{-1}H_{kl}^{-1}H_{mn}^{-1}\dfrac{\partial^6}{\partial x_i\partial x_j\partial x_k\partial x_l\partial x_m\partial x_n}\right]\left(g^2_{x_0}u\right)(0)\\=&-\dfrac{1}{48i}\sum_{i,j,k,l,m,n=1}^{44} H^{-1}_{ij}H^{-1}_{kl}H^{-1}_{mn}\left[\dfrac{\partial^3 g_{x_0}(0)}{\partial x_j\partial x_k\partial x_l}\dfrac{\partial^3 g_{x_0}(0)}{\partial x_i\partial x_m\partial x_n}\right.\\+&\dfrac{\partial^3 g_{x_0}(0)}{\partial x_j\partial x_k\partial x_m}\dfrac{\partial^3 g_{x_0}(0)}{\partial x_i\partial x_l\partial x_n}+\dfrac{\partial^3 g_{x_0}(0)}{\partial x_j\partial x_k\partial x_n}\dfrac{\partial^3 g_{x_0}(0)}{\partial x_i\partial x_l\partial x_m}\\+&\dfrac{\partial^3 g_{x_0}(0)}{\partial x_j\partial x_l\partial x_m}\dfrac{\partial^3 g_{x_0}(0)}{\partial x_i\partial x_k\partial x_n}+\dfrac{\partial^3 g_{x_0}(0)}{\partial x_j\partial x_l\partial x_n}\dfrac{\partial^3 g_{x_0}(0)}{\partial x_i\partial x_k\partial x_m}\\+&\dfrac{\partial^3 g_{x_0}(0)}{\partial x_j\partial x_m\partial x_n}\dfrac{\partial^3 g_{x_0}(0)}{\partial x_i\partial x_k\partial x_l}+\dfrac{\partial^3 g_{x_0}(0)}{\partial x_k\partial x_l\partial x_m}\dfrac{\partial^3 g_{x_0}(0)}{\partial x_i\partial x_j\partial x_n}\\+&\dfrac{\partial^3 g_{x_0}(0)}{\partial x_k\partial x_l\partial x_n}\dfrac{\partial^3 g_{x_0}(0)}{\partial x_i\partial x_j\partial x_m}+\dfrac{\partial^3 g_{x_0}(0)}{\partial x_k\partial x_m\partial x_n}\dfrac{\partial^3 g_{x_0}(0)}{\partial x_i\partial x_j\partial x_l}\\+&\left.\dfrac{\partial^3 g_{x_0}(0)}{\partial x_l\partial x_m\partial x_n}\dfrac{\partial^3 g_{x_0}(0)}{\partial x_i\partial x_j\partial x_k}\right]u\left(x_0\right).\
\end{aligned}\label{I3}
\end{equation}
We have deduced sixth-order derivative to third-order derivatives with the condition $g_{x_0}(0)=0,\ g'_{x_0}(0)=0$  and $g''_{x_0}(0)=0$. Factors in each term in the square-bracket above are elements which we have stored in table $G^{(3)}$. We use similar techniques in 2) to simplify the summation above to make the computation less expensive.
\end{itemize}
\subsection{Numerical results}
The asymptotic results for the integral (\ref{explicitamp}) are:
\begin{equation}
\begin{aligned}
A_v=& A_v^{(+)}+A_v^{(-)},\\
A^{(\pm)}_v=& A^{(\pm)0}_v+A_v^{(\pm)1}+O\left(\frac{1}{\lambda^2}\right),\\
A^{(\pm)0}_v =& 2^4 \prod_{a<b} \frac{d_{\l j_{ab}}}{\pi} e^{i\lambda \tilde{S}(\vec{\bf x}^\pm_0)} \left[\text{det}\left(\dfrac{\lambda S''(\vec{\bf x}^\pm_0)}{2\pi i}\right)\right]^{-\frac{1}{2}}u(\vec{\bf x}^\pm_0),\\
A^{(\pm)1}_v =& 2^4\prod_{a<b} \frac{d_{\l j_{ab}}}{\pi} e^{i\lambda \tilde{S}(\vec{\bf x}^\pm_0)} \left[\text{det}\left(\dfrac{\lambda S''(\vec{\bf x}^\pm_0)}{2\pi i}\right)\right]^{-\frac{1}{2}}\dfrac{1}{\lambda}\left(I_1+I_2+I_3\right)(\vec{\bf x}^\pm_0).\label{asymp}
\end{aligned}
\end{equation}
where $A_v^{(+)}$ and $A_v^{(-)}$ are asymptotic expansions of $A_v$ at two distinct critical points $\vec{\bf x}^\pm_0$ corresponding to the geometrical 4-simplex with opposite orientations. $A^{(\pm)0}_v$ stands for the leading-order term of the asymptotics and $A^{(\pm)1}_v$ is the next-to-leading order correction. The additional factor $2^{4}$ comes from the double multiplicity of the solutions $g_a=\pm g^0$ for $a\neq 1$.  We evaluate the leading-order term and the next-to-leading-order corrections at $\gamma=0.1$ as an example:
\begin{equation}
\begin{aligned}
A_v^{(+)}&=\left(1+\frac{1}{4 \lambda }\right)^6\left(1+\frac{1}{10 \lambda }\right)^4\frac{1.77\times 10^{-13}+1.87\times 10^{-14} i}{\lambda^{12}} e^{4.60\lambda i}\left(1-\frac{3.082+0.601 i}{\lambda}\right),\\
A_v^{(-)}&=\left(1+\frac{1}{4 \lambda }\right)^6\left(1+\frac{1}{10 \lambda }\right)^4\frac{1.77\times 10^{-13}-1.87\times 10^{-14} i}{\lambda^{12}} e^{4.58\lambda i}\left(1-\frac{3.082-0.601 i}{\lambda}\right),
\end{aligned}\label{Avpm}
\end{equation}
where $4.60$ and $4.58$ in exponents are values of $\tilde{S}(\vec{\bf x}_0^\pm)$. The factor \\$\left(1+\frac{1}{4 \lambda }\right)^6\left(1+\frac{1}{10 \lambda }\right)^4=\prod_{a<b}d_{\l j_{ab}}/(2\l j_{ab})$ and
\be
2^4\prod_{a<b}\frac{2\l j_{ab}}{\pi}  \left[\text{det}\left(\dfrac{\lambda S''(\vec{\bf x}^\pm_0)}{2\pi i}\right)\right]^{-\frac{1}{2}}u(\vec{\bf x}^\pm_0)=\frac{1.77\times 10^{-13}\pm 1.87\times 10^{-14} i}{\lambda^{12}}.
\ee

We obtain the asymptotics of the EPRL 4-simplex amplitude $A_v$ with the next-to-leading-order corrections:
\be
A_v&=&\left(1+\frac{1}{4 \lambda }\right)^6\left(1+\frac{1}{10 \lambda }\right)^4\frac{3.55 \times 10^{-13}}{\lambda^{12}}\,e^{4.59 \lambda i}\nonumber\\
&&\left[\cos (0.106+0.01 \lambda)+\frac{3.14}{\lambda} \sin (-1.27+0.01 \lambda)+O\lt(\frac{1}{\l^2}\rt)\right].\label{result}
\ee 
It has been arranged in terms of cosines and sines, similar to the case of 6j symbol \cite{Bonzom:2008xd}. The asymptotic amplitudes $A_v$ with the other boundary geometries can be found in Eq.(\ref{Av3and8}) and Eq.(\ref{Av4and11}). The Regge action $S_{Regge}$ of the geometrical 4-simplex is inside the cosine and sine 
\be
S_{Regge}=\l\sum_{a<b}\g j_{ab}\theta^L_{ab} =0.01\l.
\ee 
Next, we combine cosines and sines in Eq.(\ref{result}) to be in a nicer form. As $\l\to\infty$, we can expand $\log\left(1-\frac{3.082\pm 0.601i}{\lambda}\right)$ to 1st order in $1/\l$
\be
1-\frac{3.082\pm 0.601 i}{\lambda}\approx \exp\left(-\frac{3.082}{\lambda}\right)\exp\left({\pm\frac{0.601i}{\lambda}}\right).
\ee 
Hence, (\ref{result}) can be rewritten as
\be
A_v&\simeq&\frac{1}{\lambda^{12}}  \lt( e^{S^{(+)}_{eff}}+e^{S^{(-)}_{eff}}\rt).\label{result2}
\ee 
where 
\be
S^{(\pm)}_{eff}=\pm i\lt(0.01\lambda+0.106-\frac{0.601}{\lambda}\rt)+4.59 \lambda i-28.6667 - \frac{1.182}{\l}
\ee
can be viewed as a ``quantum effective action'' providing quantum correction to the Regge action\footnote{In the functional integration of QFT, $e^{-i E[J]}=\int \mathcal{D} \phi e^{\frac{i}{\hbar} \int d^{4} x\left(\mathcal{L}[\phi]+J \phi\right)}$ where $J$ is the source. When we expand the path integral at a classical solution (a critical point) $\phi_{\rm cl}$, we obtain (see P.372 in \cite{Peskin:257493})
\be
-i E[J]=\frac{i}{\hbar} \int d^{4} x\left(\mathcal{L}\left[\phi_{\mathrm{cl}}\right]+J \phi_{\mathrm{cl}}\right)-\frac{1}{2} \log \operatorname{det}\left[-\frac{\delta^{2} \mathcal{L}}{\delta \phi \delta \phi}\right] + O(\hbar)
\ee
where $O(\hbar)$ correponds to re-exponentiaing $1+O(1/\l)\sim e^{O(1/\l)}$ from Theorem \ref{hormanderThm}. In our case, $O(1/{\lambda})$ is the analog of $O(\hbar)$. The quantum effective action $\Gamma\left[\phi_{\mathrm{cl}}\right]$ is defined by the Legendre transformation $\Gamma\left[\phi_{\mathrm{cl}}\right] \equiv-E[J]-\int d^{4} x J \phi_{\mathrm{cl}}$. The quantum effective action evaluated at $\phi_{\rm cl}$ is given by
\be
\Gamma\left[\phi_{\mathrm{cl}}\right]=\frac{i}{\hbar} \int d^{4} x\mathcal{L}\left[\phi_{\mathrm{cl}}\right]-\frac{1}{2} \log \operatorname{det}\left[-\frac{\delta^{2} \mathcal{L}}{\delta \phi \delta \phi}\right] + O(\hbar)
\ee
where $O(\hbar)$ is the same as the one in $E[J]$ in the perturbative expansion. Our $S^{(\pm)}_{eff}$ is an analog of $\G$ when we view the spinfoam amplitude as an analog of the functional integration. $S^{(\pm)}_{eff}$ gives $1/\l$ correction to the leading Regge action in the same way as $\G$ gives $O(\hbar)$ correction to the classical action $\int d^{4} x\cl$. Here we have both $S^{(\pm)}_{eff}$ since there are 2 critical points.}. The term $4.59 \lambda i$ is from the overall phase.

\begin{figure}[t]	
	\begin{center}
	\includegraphics[width=1\linewidth]{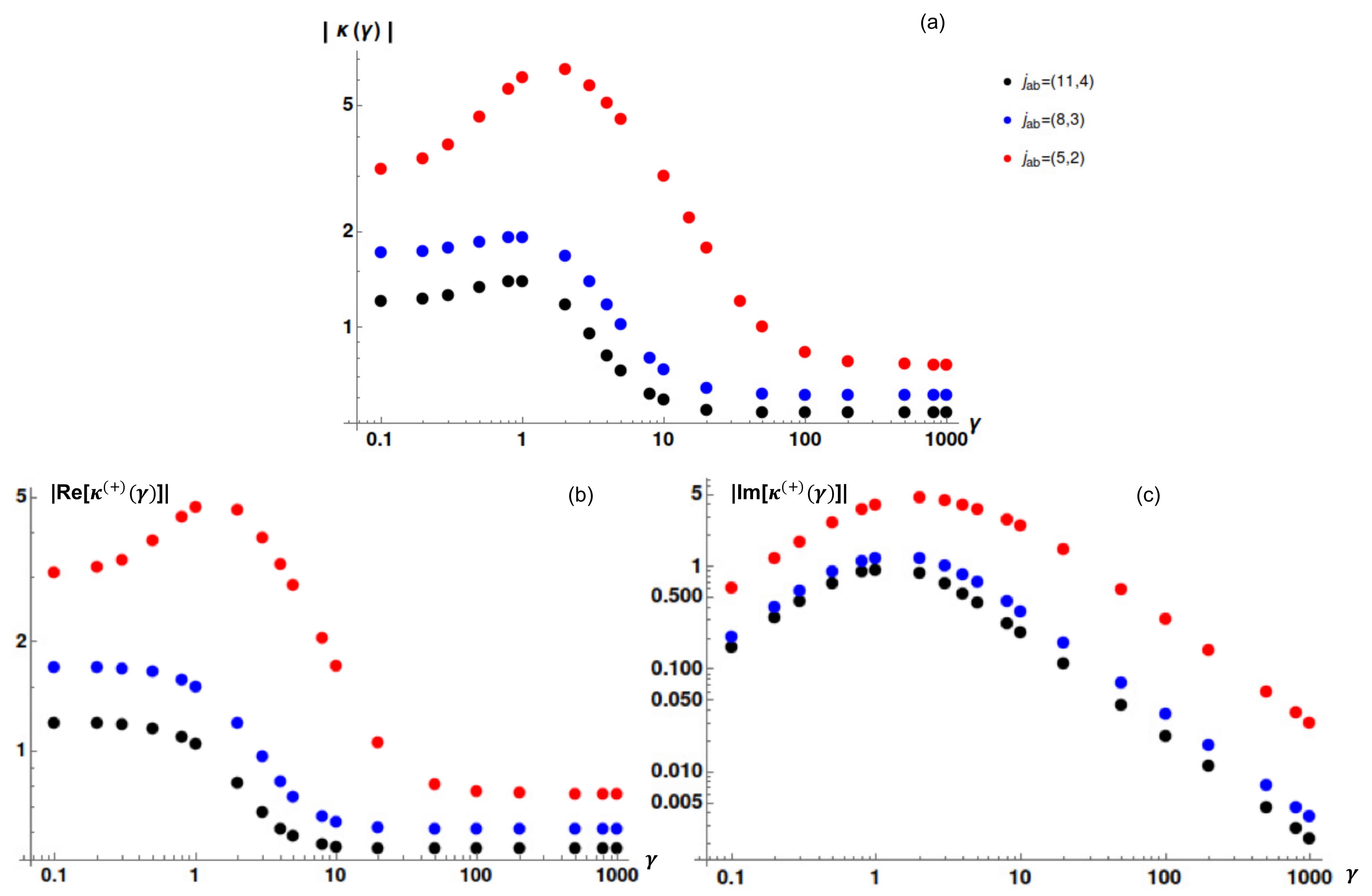}
	\caption{(a). A log-log plot of the specified list of $\gamma$ and $\left|\kappa(\gamma)\right|$ values. (b). List log-log plot of $\gamma$ and the absolute value of real part of $\kappa^{(+)}(\gamma)$. (c). List log-log plot of $\gamma$ and the absolute value of the imaginary part of $\kappa^{(+)}(\gamma)$. We use different colors to denote various boundary geometries $j_{ab}=(5,2)$ (red solid points), $j_{ab}=(8,3)$ (blue solid points), $j_{ab}=(11,4)$ (black solid points).}
	\label{asympgraph}
	\end{center}
\end{figure}

Both the leading-order term and the next-to-leading-order corrections depend on $\gamma$, we write the asymptotic amplitude as 
\be
A_v^{(\pm)}\approx C^{(\pm)}(\gamma) \left(1+\frac{\kappa^{(\pm)}(\gamma)}{\lambda}\right),\label{kappa}
\ee  
which reduces to (\ref{Avpm}) when $\g=0.1$. Here, $C^{(\pm)}(\gamma)$ coincides with the leading-order asymptotics. $\kappa^{(\pm)}(\g)$ is the next-to-leading order coefficient. $C^{(\pm)}(\gamma), \kappa^{(\pm)}$ depending on the value of $\g$. The ratio of the next-to-leading order corrections to the leading-order term is
\be
\frac{\kappa^{(\pm)}(\gamma)}{\lambda}=\frac{1}{\lambda}\frac{I^{(\pm)}(\gamma)}{u(\vec{\bf x}_0^\pm)},
\ee
here, $u(\vec{\bf x}_0^+)=\frac{0.0073}{2^{48}\pi^{16}}$ and $u(\vec{\bf x}_0^-)=\frac{0.28}{2^{48}\pi^{16}}$ are independent on $\gamma$, $I^{(\pm)}(\gamma)=(I_1+I_2+I_3)(\vec{\bf x}_0^\pm)$ depends on $\gamma$, and 
\be
\kappa^{(+)}(\g)=\overline{\kappa^{(-)}(\g)}.
\ee
We use $\left|\kappa(\gamma)\right|$ to denote $\left|\kappa^{(+)}(\gamma)\right|$ and $\left|\kappa^{(-)}(\gamma)\right|$. 
We show some results of $|I^{(\pm)}(\gamma)|$ and $\left|\kappa(\gamma)\right|$ at different $\g$ in Table \ref{tab:correction}. We plot $\left|\kappa(\gamma)\right|$ and $\kappa^{(+)}(\gamma)$ versus $\gamma$ in Figure \ref{asympgraph}.

\begin{table}[H]
	\small
	\centering\caption{$\left|I^{(\pm)}(\gamma)\right|$ and $\left|\kappa(\gamma)\right|$ with respect to $\gamma$. 
	}\label{tab:correction}
	\begin{tabular}{|c|c|c|c|c|c|c|c|c|}
		\hline
		$\gamma$&0.1&0.5&1&2&3&4&5&8\\
		\hline
		$2^{48}\pi^{16}\left|I^{(+)}(\gamma)\right|$&$0.023$&$0.033$&$0.044$&$0.047$&$0.042$&$0.037$&$0.033$&$0.025$\\
		\hline
		$2^{48}\pi^{16}\left|I^{(-)}(\gamma)\right|$&$0.87$&$1.28$&$1.70$&$1.80$&$1.60$&$1.41$&$1.26$&$0.97$\\
		\hline
		$\left|\kappa(\gamma)\right|$&$3.14$&$4.6$&$6.10$&$6.46$&$5.75$&$5.07$&$4.53$&$3.46$\\
		\hline
		$\gamma$&10&20&50&100&200&500&800&1000 \\
		\hline
		$2^{48}\pi^{16}\left|I^{(+)}(\gamma)\right|$&$0.022$&$0.013$&$0.007$&$0.006$&$0.0056$&$0.0055$&$0.0055$&$0.0055$\\
		\hline
	    $2^{48}\pi^{16}\left|I^{(-)}(\gamma)\right|$&$0.83$&$0.50$&$0.28$&$0.23$&$0.22$&$0.21$&$0.21$&$0.21$\\
		\hline
		$\left|\kappa(\gamma)\right|$&$2.99$&$1.77$&$1.01$&$0.83$&$0.78$&$0.77$&$0.76$&$0.76$\\
		\hline
	\end{tabular}
\end{table}

Table \ref{tab:correction} and Figure \ref{asympgraph} demonstrate how the next-to-leading order corrections change with different values of $\gamma$. For the specific boundary geometry $j_{ab}=(5,2)\l$, $\left|\kappa(\gamma)\right|$ increases first and then decrease with increasing $\gamma$, and the maximum occurs at about $\gamma=2$. The red points on Figure \ref{asympgraph} (b) and (c) show that $\kappa^{(\pm)}(\gamma)$ stabilize to real constants $0.76$ as $\gamma\rightarrow\infty$. We also perform the analysis to other boundary geometries $j_{ab}=(11,4)\l$ and $j_{ab}=(8,3)\l$ (black and blue points in Figure \ref{asympgraph}). In the case of $j_{ab}=(8,3)\l$, $\kappa^{(\pm)}(\gamma)$ stabilizes to real constants $0.61$ as $\gamma\rightarrow\infty$. For $j_{ab}=(11,4)\l$, $\kappa^{(\pm)}(\gamma)$ maintain at real constants $0.54$ as $\gamma\rightarrow\infty$. We can conclude that $\kappa^{(\pm)}(\gamma)$ will stabilize to a real constant as $\gamma\rightarrow\infty$.

The semiclassical approximation of $A_v$ with the leading order as in \cite{Barrett:2009mw} is valid when $\l$ is large enough so that $\frac{\kappa^{(\pm)}(\gamma)}{\lambda}$ can be negligible comparing to 1. For example, at $\lambda=30$ and $\gamma=0.1$, $\left|\kappa(\gamma)\right|/\l\approx 0.10$ is about 10\% of the leading order. In our opinion, a much safer regime for validating the semiclassical approximation needs an even larger $\l$, e.g., $\l=300$ so that $\left|\kappa(\gamma)\right|/\l\approx 0.01$.

\section{Next-to-leading order correction in large-$j$ 4-simplex amplitude with coherent spin-network boundary state}\label{spinnetwork}

\subsection{Coherent spin-network state}

Above discussions take the coherent intertwiners as the boundary state for the 4-simplex amplitude. In this section, we use the coherent spin-networks as the boundary state. The coherent spin-networks relate to coherent intertwiners by a superposition
\be
\left|\Psi_0\right\rangle=\sum_{j_{ab}\in\Z_+/2\cup\{0\}}\psi_{j_0,\phi_0}(\vec{j})\otimes_{a=1}^5|i_a(\vec{j},\vec{\xi}) \rangle,\label{semi-cohere}
\ee
where $\psi_{j_0,\phi_0}(\vec{j})$ is given by,
\begin{equation}
	\psi_{j_0, \phi_0}(\vec{j})=\exp \left(-i \sum_{a b} \gamma \phi_0^{a b}\left(j_{a b}-\left(j_0\right)_{a b}\right)\right)
	 \exp \left(-\half\sum_{a b, c d} \alpha^{(a b)(c d)} \frac{j_{a b}-\left(j_0\right)_{a b}}{\sqrt{\left(j_0\right)_{a b}}} \frac{j_{c d}-\left(j_0\right)_{c d}}{\sqrt{\left(j_0\right)_{c d}}}\right),
\end{equation}
which is a Gaussian times a phase. $(j_0)_{ab}\in\Z_+/2$. $\phi_0^{ab}$ is the discrete extrinsic curvature relating to the dihedral angle of the triangle $ab$ in a 4-simplex geometry. Here we first choose $(j_0)_{ab}=(5,2)\l $ and $\xi_{ab}$ are the same boundary data as the above discussions. We set values of $\phi_0^{ab}$ by 
\be
\g \phi_0^{ab}=\frac{\partial \tilde{S}(\vec{j}, \vec{\bf x}_0^+)}{\partial j_{ab}}.\label{setupphi0}
\ee
$\alpha^{\left(ab\right)\left(cd\right)}$ is a $10\times 10$ matrix given by
\be
\alpha^{(a b)(c d)}=\alpha_{1} \delta^{(a b)(c d)}+\alpha_{2} m^{(a b)(c d)}+\alpha_{3} n^{(a b)(c d)},
\ee
$\delta^{(a b)(c d)}=1$ if $(a b)=(c d)$, $m^{(a b)(c d)}=1$ if just two indexes are the same, and $n^{(a b)(c d)}=1$ if all four indexes are different, and in all other cases these quantities vanish \cite{Rovelli:2005yj}. The coherent spin-networks depend on 3 free parameters $\alpha_1,\alpha_2,\alpha_3$ \footnote{\cite{Bianchi:2009ky} shows that coherent spin-networks with $\a_2=\a_3=0$ relates to Thiemann's coherent state \cite{Thiemann:2000bw}.}. We choose $\alpha_1=2,\alpha_2=3,\alpha_3=4$ in our computation. The spinfoam amplitude with coherent spin-networks as the boundary state depends on the choice of $\alpha$'s. Our result depends on our choice of values of $\alpha$'s. 

The EPRL 4-simplex amplitude with coherent spin-network boundary sums $A_v$ over $j_{ab}$ weighted by $\psi_{j_0,\phi_0}(\vec{j})$. 
The 4-simplex amplitude for a coherent spin network state 
\begin{equation}
\begin{aligned}
A'_v&=\sum_{j_{ab}\in\Z_+/2\cup\{0\}} \psi_{j_0,\phi_0}(\vec{j}) A_v(j_{ab},i_a)  \\
&= \sum_{j_{ab}\in\Z_+/2\cup\{0\}}\int \prod_{a=2}^{5} dg_a \int_{(\mathbb{CP}^1)^{10}} e^{S_{\mathrm{tot}}} \prod_{a<b}\frac{d_{j_{ab}}}{\pi}\Omega_{ab}, \label{amptot}
\end{aligned}
\end{equation}
where the "total action" $S_{\rm tot}$ is given by 
\begin{equation}
S_{\mathrm{tot}}(j_{ab},g,\mathbf{z})=-\frac{1}{2} \sum_{a b, c d} \alpha^{(a b)(c d)} \frac{j_{a b}-\left(j_0\right)_{a b}}{\sqrt{\left(j_0\right)_{a b}}} \frac{j_{c d}-\left(j_0\right)_{c d}}{\sqrt{\left(j_0\right)_{c d}}}-i \sum_{a b} \gamma \phi_0^{a b}\left(j_{a b}-\left(j_0\right)_{a b}\right)+S(j, g, \mathbf{z}),\label{Stot}
\end{equation}
where $S$ is the same action as (\ref{action0}). 

We use Poisson re-summation formula
\be
\sum_{j\in Z_+/2\cup\{0\}}f(j)&=&\half \sum_{j\in\Z/2}f(|j|) +\half f(0)\\ &=& 2\sum_{k\in\Z} \int_0^\infty dj\ f(j)\, e^{4\pi i k j}+\half f(0)
\ee
where $f(j)$ corresponds to the summand in Eq.(\ref{amptot}). When $(j_0)_{ab}$ are large, the Gaussian in $\psi_{j_0,\phi_0}$ is peaked at large spins $j_{ab}= (j_0)_{ab}$. $|\psi_{j_0,\phi_0}|$ is exponentially small when $j_{ab}$ is far from the large $(j_0)_{ab}$, so $f(0)$ is exponentially small and negligible. Similarly, the integral $\int dj $ is dominated by the large-$j$ domain with $j_{ab}\sim (j_0)_{ab}$, while the integral outside this domain is exponentially suppressed. Motivated by this, we scale $j_{ab}$ and $(j_0)_{ab}$ by $j_{ab}\rightarrow\lambda j_{ab}$ and $(j_0)_{ab}\rightarrow\lambda(j_0)_{ab}$. Therefore, ``total action" is scaled by $S_{\mathrm{tot}}\rightarrow\lambda S_{\mathrm{tot}}$, and 
\be
A'_v=(2\l)^{10}\sum_{k_{ab}\in\Z}\int\prod_{a<b}dj_{ab}\frac{d_{\l j_{ab}}}{\pi}\int \prod_{a=2}^{5} dg_a \int_{(\mathbb{CP}^1)^{10}} e^{\l S^{(k)}_{\mathrm{tot}}} \prod_{a<b}\Omega_{ab}\label{Aprimvsumk}
\ee
where
\be
S^{(k)}_{\mathrm{tot}}=S_{\mathrm{tot}}+4\pi i \sum_{a<b}j_{ab}k_{ab}
\ee
Integrals in Eq.(\ref{Aprimvsumk}) can be analyzed with stationary phase approximation as in Theorem \ref{hormanderThm}. Critical point equations of each $S^{(k)}_{\rm tot}$ are
\be
\mathrm{Re}(S_{\rm tot})=0,\quad \partial_{j_{ab}}S_{\rm tot}=4\pi i k_{ab},\quad \partial_{g_a}S= \partial_{z_{ab}}S=0.\label{criticalAprime}
\ee
It is not hard to see these equations imply critical equations of $S$ in Eqs.(\ref{eq:criticalpt}) and $j_{ab}=(j_0)_{ab}$. Among two solutions $\vec{\bf x}^{(\pm)}_0$ of Eq.(\ref{eq:criticalpt}), only $\vec{\bf x}^{(+)}_0$ satisfy Eq.(\ref{criticalAprime}) when all $k_{ab}=0$. Any $k_{ab}\neq 0$ leads to no solution for Eqs.(\ref{setupphi0}). Therefore all integrals except for all $k_{ab}=0$ in (\ref{Aprimvsumk}) are suppressed as $O(\l^{-N})$ for all positive integer $N$.

We focus on all $k_{ab}=0$ and neglect exponentially small errors
\begin{equation}
\begin{aligned}
	A'_v&=(2\l)^{10}\int \prod_{a<b} dj_{ab}\lt(\frac{d_{\l j_{ab}}}{\pi}\rt) \int \prod_{a=2}^{5} dg_a  \int_{(\mathbb{CP}^1)^{10}} e^{\lambda S_{\mathrm{tot}}} \prod_{a<b}\Omega_{ab}\\
	&=(2\l)^{10}\int \prod_{i=1}^{54} d \eta_{i}\, u'({\bm \eta}) e^{i \lambda \tilde{S}_{\mathrm{tot}}({\bm \eta})},\quad {\bm \eta}=\lt(\{j_{ab}-(j_0)_{ab}\}_{a<b},\vec{\bf x}\,\rt)\label{asymStot}
\end{aligned}
\end{equation}
where $\tilde{S}_{\mathrm{tot}}=-i S_{\mathrm{tot}}$ and $u'({\bm \eta})=u(\vec{\bf x})\prod_{a<b}({d_{\l j_{ab}}}/{\pi})$. The asymptotic expansion (\ref{theorem}) can be applied to compute (\ref{asymStot}). $A'_v$ has only one critical point given by $j_{ab}=(j_0)_{ab}$ and $\vec{\bf x}=\vec{\bf x}_0^+$, because the boundary coherent spin-network specifies both boundary 3-geometry and extrinsic curvature \cite{Bianchi:2010mw}.

\subsection{Numerical results}

The asymptotic expansion of $A'_v$ with the next-to-leading order correction can be computed with the same scheme as Section \ref{Asymptotic Expansion with the coherent-intertwiner boundary state}. Numerical results are presented below.

As an example, at $\gamma=0.1$, 
\be
	A'_v &=&C'(\gamma) \left[1+\frac{\kappa'(\gamma)}{\lambda}+O\lt(\frac{1}{\l^2}\rt)\right]\\
	&=&2^{10}\,\left(1+\frac{1}{4 \lambda }\right)^6\left(1+\frac{1}{10 \lambda }\right)^4\frac{1.85\times 10^{-9}+9.31\times 10^{-10} i}{\lambda^{7}}\,\e^{4.60 \lambda i }\nonumber\\
	&&\left[1+\frac{26.58+30.78 i}{\lambda}+O\left(\frac{1}{\l^2}\right)\right],
\label{nextResult}
\ee
where $C'(\gamma)$ stands for the leading-order terms. $\kappa'(\g)$ is the next-to-leading order coefficient. The asymptotic amplitudes $A'_v$ with the other boundary geometries can be found in Eq.(\ref{coh3and8}) and Eq.(\ref{coh4and11}). The ratio of the next-to-leading order corrections to the leading-order terms is
\be
\frac{\kappa'(\gamma)}{\lambda}=\frac{1}{\lambda}\frac{I'(\gamma)}{u'(0)},
\ee
here, $I'=I'_1+I'_2+I'_3$ is obtained by applying the computation in Eq.(\ref{I1}), (\ref{I2}) and (\ref{I3}) to $u'$ and $\tilde{S}_{\rm tot}$. $u'(0)=\frac{3.18}{2^{48}\pi^{26}}$ in the leading-order term is independent of $\gamma$.

From the result of Eq.(\ref{nextResult}), the next-to-leading order coefficient gives $|\kappa'(0.1)|\simeq 40.67$ at $\g=0.1$. When $\l=30$, $|{\kappa'(0.1)}/{\lambda}|\simeq 1.36$ is even larger than the leading-order term. Clearly, the expansion in this case is invalid at $\l=30$. The semiclassical approximation of $A'_{v}$ (approximation by the leading order) requires a much larger $\l$. For example, $\l\geq 300$, then $|{\kappa'(0.1)}/{\lambda}|$ is bounded by about 13\% of the leading order. We suggest $\l\geq 3000$ to be a much better regime for $A'_v$ at $\g=0.1$ where the next-to-leading order term is about $1\%$ of the leading order.

\begin{figure}[H]	
	\begin{center}
		\includegraphics[width=0.85\linewidth]{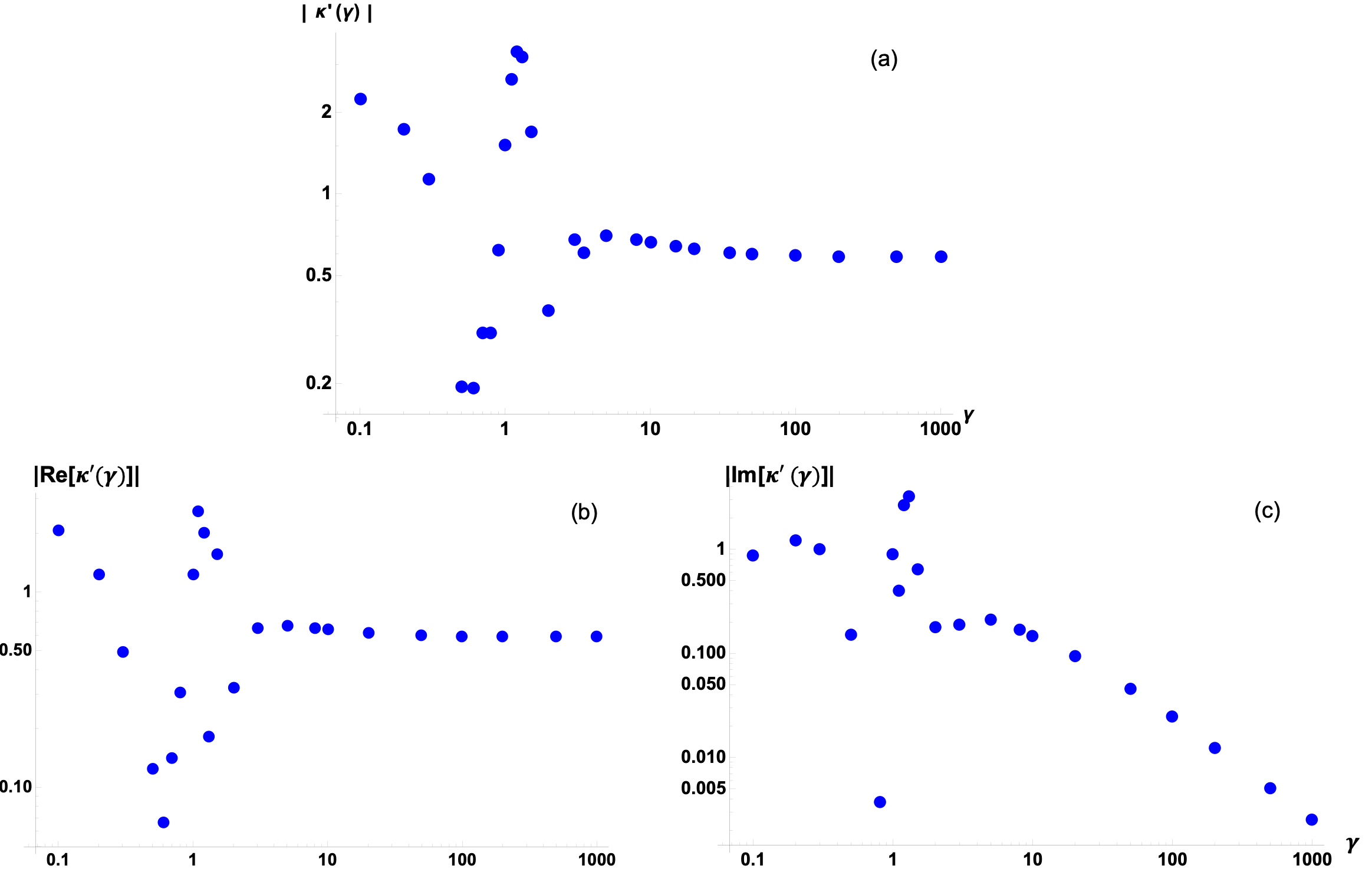}
		\caption{The numerical results with boundary geometry $j_{ab}=(8,3)\l$. (a): the log-log plot of $\left|\kappa'(\gamma)\right|$ versus $\g$. (b): the log-log plot of the absolute value of the real part of $\kappa'(\gamma)$. (c): the log-log plot of absolute value of the imaginary part of $\kappa'(\gamma)$.}
		\label{coh8and3}
	\end{center}
\end{figure}

\begin{figure}[H]	
	\begin{center}
		\includegraphics[width=0.85\linewidth]{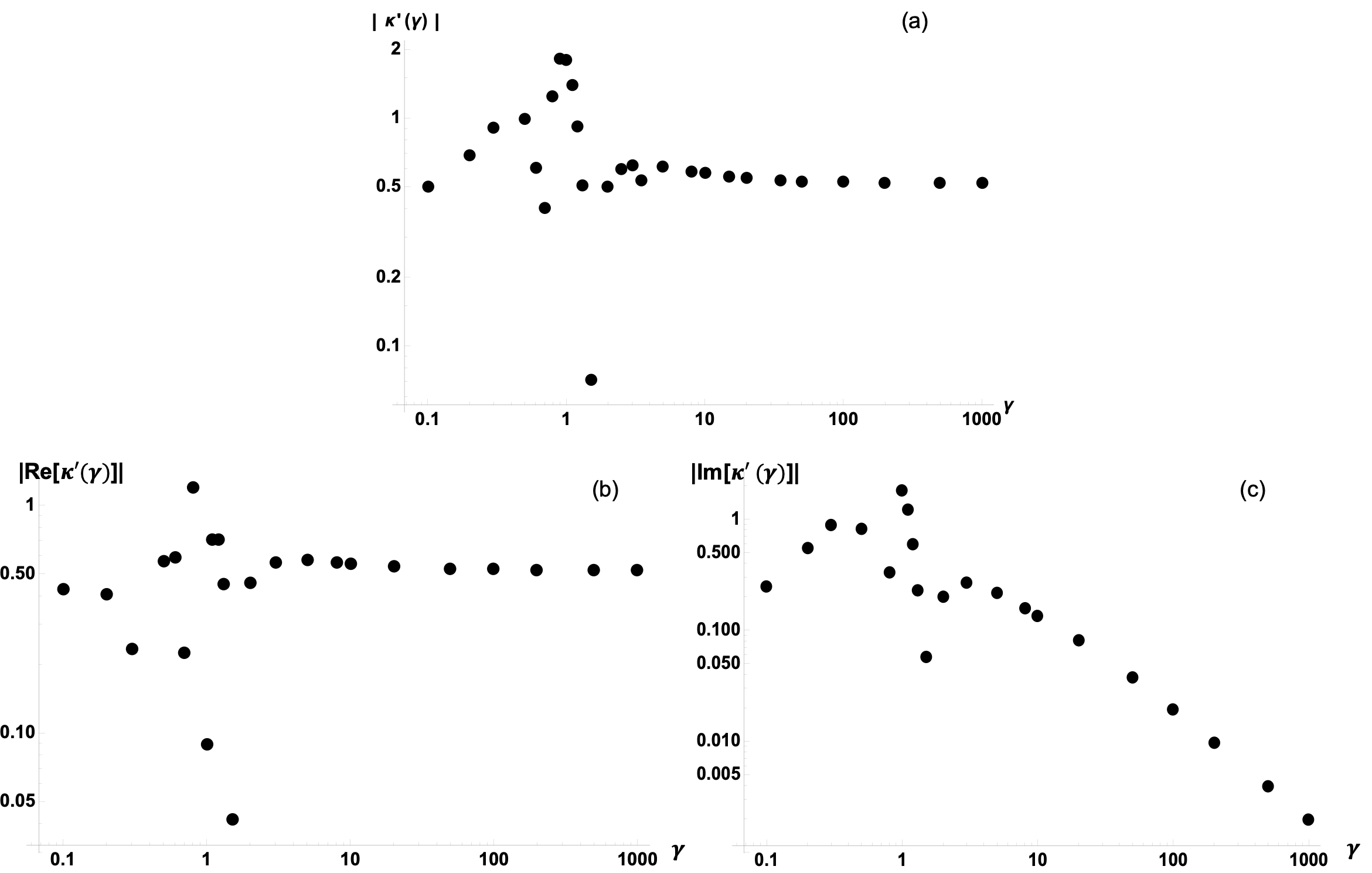}
		\caption{The numerical results with boundary geometry $j_{ab}=(11,4)\l$. (a): the log-log plot of $\left|\kappa'(\gamma)\right|$ versus $\g$. (b): the log-log plot of the absolute value of the real part of $\kappa'(\gamma)$. (c): the log-log plot of absolute value of the imaginary part of $\kappa'(\gamma)$.}
		\label{coh11and4}
	\end{center}
\end{figure}

This increase is not universal but only happens in certain examples. There are other examples (with different boundary geometries) giving $|\kappa'(0.1)|\sim O(1)$ and not requiring a large increase of $\l$. For examples, $|\kappa'(0.1)|\simeq 0.50$ when $j_{ab}=(11,4)\l$, and $|\kappa'(0.1)|\simeq 2.23$ when $j_{ab}=(8,3)\l$, see Fig.\ref{coh11and4} and Fig.\ref{coh8and3}. It illustrates that $|\kappa'(0.1)|$ damps off significantly as $j_{ab}/\l$ goes from $(5,2)$ to $(11,4)$. Our numerical studies demonstrate these results, although we don't have a mathematical argument for the increase/non-increase of $\l$ due to the complication of the sum at the next-to-leading order corrections.

\begin{figure}[H]	
	\begin{center}
		\includegraphics[width=0.85\linewidth]{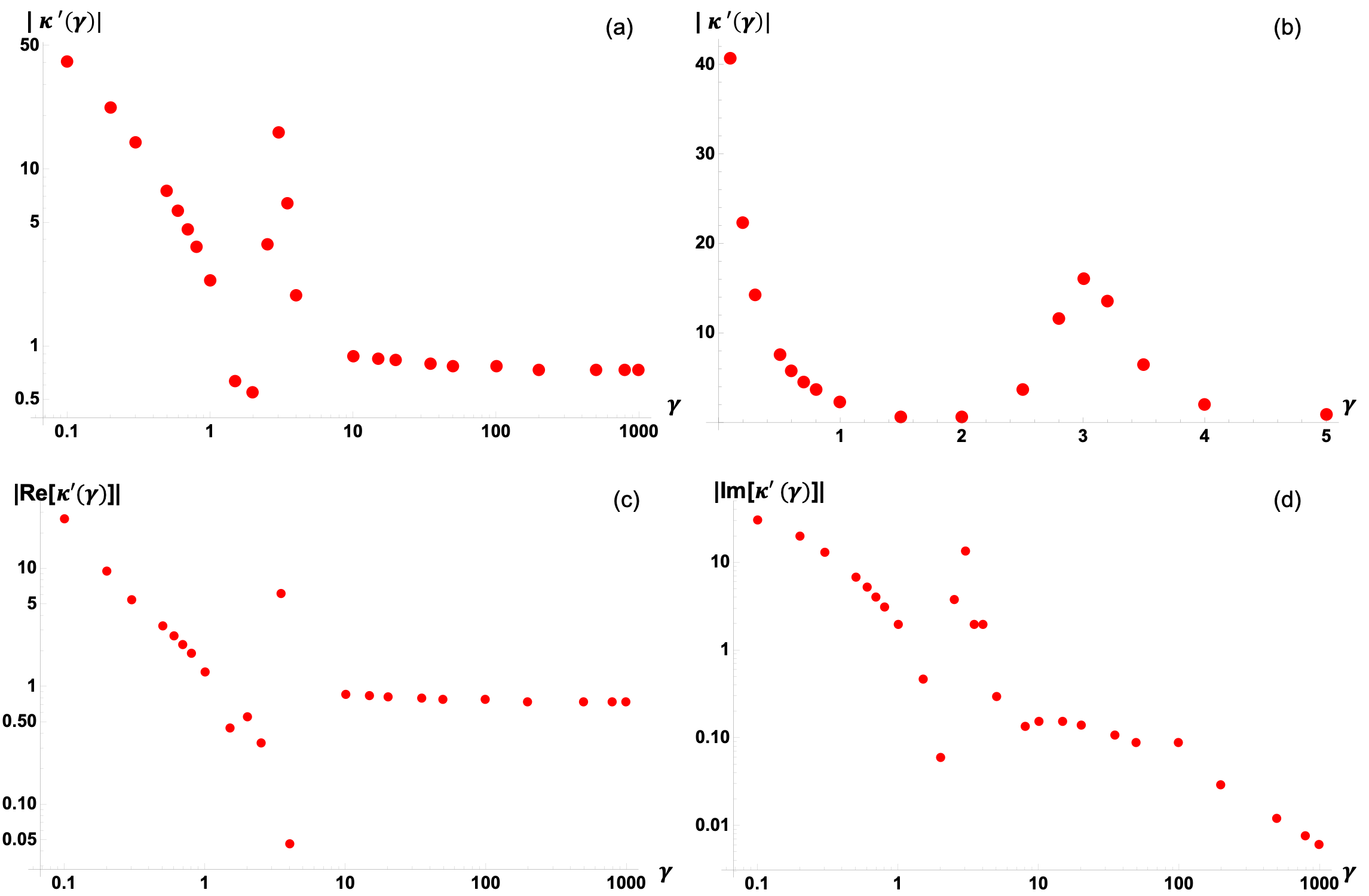}
		\caption{The numerical results with boundary geometry $j_{ab}=(5,2)\l$. (a): the log-log plot of $\left|\kappa'(\gamma)\right|$ versus $\g$. (b): the plot of $\left|\kappa'(\gamma)\right|$ with relatively small $\gamma$. Panel (b) is a zoom of panel (a) for $\gamma\in\lbrack0.1,5\rbrack$. (c): the log-log plot of the absolute value of the real part of $\kappa'(\gamma)$. (d): the log-log plot of absolute value of the imaginary part of $\kappa'(\gamma)$.}
		\label{coh5and2}
	\end{center}
\end{figure}

Moreover, we study numerically the dependence of $\kappa'$ on $\g$. We list some results of $\left|I'(\gamma)\right|$ and $\left|\kappa'(\gamma)\right|$ at different values of $\gamma$ in Table \ref{tab1} for $(j_0)_{ab}=(5,2)\l$. The plot of $\left|\kappa'(\gamma)\right|$ versus $\gamma$ is given by Fig.\ref{coh5and2}(a), Fig.\ref{coh8and3}(a) and Fig.\ref{coh11and4} (a).

\begin{table}[H]
	\small
	\centering\caption{$\left|I'(\gamma)\right|$ and $\left|\kappa'(\gamma)\right|$ at different $\gamma$. 
	}\label{tab1}
	\begin{tabular}{|c|c|c|c|c|c|c|c|c|}
		\hline
		$\gamma$&0.1&0.5&1&2&3&4&5&8\\
		\hline
		$2^{48}\pi^{26}\left|I'(\gamma)\right|$&$129.13$&$23.86$&$7.45$&$1.75$&$50.96$&$6.17$&$2.91$&$2.77$\\
		\hline
		$\left|\kappa'(\gamma)\right|$&$40.67$&$7.51$&$2.34$&$0.55$&$16.05$&$1.94$&$0.92$ & $0.87$\\
		\hline
		$\gamma$&10&20&50&100&200&500&800&1000\\
		\hline
		$2^{48}\pi^{26}\left|I'(\gamma)\right|$&$2.77$&$2.64$&$2.46$&$2.34$&$2.32$&$2.32$&$2.32$&$2.32$\\
		\hline
		$\left|\kappa'(\gamma)\right|$&$0.87$&$0.83$&$0.77$&$0.75$&$0.74$&$0.73$&$0.73$&$0.73$\\
		\hline
	\end{tabular}
\end{table}

In the case of boundary geometry $(j_0)_{ab}=(5,2)\l$, Fig. \ref{coh5and2} (c) and (d) indicate that $\kappa'(\gamma)$ stabilizes to a real constant 0.73 asymptotically as $\g\to\infty$. For $(j_0)_{ab}=(11,4)\l$ and $(j_0)_{ab}=(8,3)\l$, $\kappa'(\gamma)$ maintain at real constants $0.58$ and $0.52$ respectively as $\gamma\rightarrow\infty$.  From Fig. \ref{coh5and2}, Fig. \ref{coh8and3} and Fig. \ref{coh11and4}, we can conclude that the next-to-leading-order corrections depend on $\gamma$,  $\left|\kappa'(\gamma)\right|$ oscillates first for small $\gamma$, but it will stabilize to a constant as $\gamma\rightarrow\infty$. For small $\g$, $|\kappa'(\gamma)|$ is relatively large and results in that $\l$ has to be large for $\g=0.1$, $(j_0)_{ab}=(5,2)\l$, while it becomes smaller for $(j_0)_{ab}=(11,4)\l$ or $(j_0)_{ab}=(8,3)\l$.

\section{Conclusion}

In this paper, we use the coherent intertwiners and coherent spin-networks respectively as boundary states to study the large-$j$ asymptotic expansion of the EPRL 4-simplex amplitude. We numerically derive the next-to-leading order corrections and compare them to the leading-order terms. We demonstrate how the next-to-leading order corrections depend on the Barbero-Immirzi parameter $\g$, and how to obtain quantum corrections to the Regge action. In the context of this, our work makes it possible to quantitatively describe the quantum behavior of 4-simplex amplitude. The results help to estimate a proper regime defined by $\lambda$ where the semiclassical approximation of $A_v$ is valid, with a dominant leading-order term and a negligible next-to-leading order correction.

It is important to extract predictions of quantum gravity effect from the spinfoam LQG as a candidate theory of quantum gravity. Our work propose to study the spinfoam amplitude perturbatively in the large-$j$ regime and understand the quantum gravity correction to be $O(1/j)$ corrections in the EPRL spinfoam amplitude. We demonstrate that this proposal can be successfully carried out by numerical computations, at least at the level of one 4-simplex amplitude. Existing results on the large-$j$ EPRL spinfoam amplitude has only been focused on the semiclassical consistency by neglecting $O(1/j)$. Moreover, Our method of computation is a straight-forward application of the stationary phase expansion of oscillatory integral, and the same method (and the Mathematica notebooks in \cite{qudx.org}) can be adapted to any spinfoam vertex amplitude.



The future generalization of this work may be along two directions: spinfoam amplitudes with multiple 4-simplices and nonperturbative computations. The challenge of generalizing to multiple 4-simplices relates to increasing number of integration variables, which makes the computation in Section \ref{Asymptotic Expansion with the coherent-intertwiner boundary state} more expensive. However, it may be still interesting and possible to study the complex with three 4-simplices as the model in \cite{Dona:2020tvv} and understand how the next-to-leading order correction interacts with the issue of flatness in the spinfoam model.

The other direction is to numerically evaluate the spinfoam amplitude nonperturbatively (without the asymptotic expansion), in order to understand the model both in and beyond the large-$j$ regime. It has been difficult since the integral (\ref{ampaction}) is oscillatory which makes numerical evaluation difficult. However, recent developments in lattice gauge theories discover new Monte-Carlo methods on Lefschetz thimbles for oscillatory integrals \cite{Witten:2010cx,Bedaque:2017epw}. The strategy is firstly deforming the integration contour to integral cycles called Lefschetz thimbles on which $\mathrm{Im}(S)$ is a constant, then applying the Monte-Carlo simulation to non-oscillatory integrals on Lefschetz thimbles. We have applied this method to study the amplitude $A'_v$, and results will be reported elsewhere \cite{toappear}.


\section*{Acknowledgements}

The authors acknowledge Pietro Dona, Giorgio Sarno, and Simone Speziale for helpful discussions. This work receives support from the National Science Foundation through grant PHY-1912278. 

\appendix

\section{The SL$\left(2,\mathbb{C}\right)$ Haar measure}\label{Haar measure}
Here, we derive the SL$\left(2,\mathbb{C}\right)$ Haar measure $dg$ in our case. \\
For any SL$\left(2,\mathbb{C}\right)$ group element, it can be parameterized as:
\begin{equation}
		g=a_0 I+\sum_{k=1}^{3} a_k\sigma_k=\left(\begin{matrix}
	a_0+a_3&a_1- i a_2\\a_1+i a_2&a_0-a_3
	\end{matrix}\right)\\=\left(\begin{matrix}
	\alpha&\beta\\\gamma&\omega
	\end{matrix}\right)=\left(\begin{matrix}
	\alpha_1+i \alpha_2&\beta_1+ i \beta_2\\ \gamma_1+i \gamma_2&\omega_1+i \omega_2 \end{matrix}\right),\label{parameter0}
\end{equation}
here, $I$ is $2\times 2$ identity matrix, $\sigma_k$ is Pauli matrix,  and $a_i(i=0,1,2,3)$,
$$
\alpha=\alpha_1+i \alpha_2,\quad \beta=\beta_1+i \beta_2,
$$
$$\gamma=\gamma_1+i \gamma_2, \quad \omega=\omega_1+i \omega_2,$$ 
are complex variables, where $\alpha_1,\alpha_2, \beta_1,\beta_2,\gamma_1,\gamma_2,\omega_1,\omega_2$ are real.  \\
From the book \cite{Ruhl}, the measure for the group SL$\left(2,\mathbb{C}\right)$ is  
\begin{equation}
\begin{aligned}
	dg&=c_0^2\delta(a_0^2-\sum_{k=1}^{3} a_k^2-1)\prod_{i=0}^{3}Da_i,\quad c_0=\pi^{-2}\\
	&=\frac{1}{\pi^4}\delta(a_0^2-\sum_{k=1}^{3} a_k^2-1)\left|\det\left(\frac{\partial\left(Re(a_0),Im(a_0),...,Re(a_3),Im(a_3)\right)}{\partial\left(Re(\alpha_1),Im(\alpha_2),...,Re(\omega_1),Im(\omega_2)\right)}\right)\right|D\alpha D\beta D\gamma D\omega\\
	&=\frac{1}{16\pi^4}\delta(\alpha\omega-\gamma\beta-1)D\alpha D\beta D\gamma D\omega,\\
\end{aligned}\label{dg}
\end{equation}
here, we use this calculation 
$$
	a_0^2-\sum_{k=1}^{3} a_k^2-1=\alpha\omega-\gamma\beta-1,
$$
$$
\det\left(\frac{\partial\left(Re(a_0),Im(a_0),...,Re(a_3),Im(a_3)\right)}{\partial\left(Re(\alpha_1),Im(\alpha_2),...,Re(\omega_1),Im(\omega_2)\right)}\right)=\frac{1}{16}.
$$
One can find the details in our Mathematica notebooks \cite{qudx.org}.
For any complex variable $z=x+i y$, we use the notation:
$$
Dz=dxdy \quad\text{and}\quad \delta(z)=\delta(x)\delta(y).
$$
Then, (\ref{dg}) can be derived as
\begin{equation}
\begin{aligned}
dg&=\frac{1}{16\pi^4}\delta(\alpha_1\omega_1-\alpha_2\omega_2-\beta_1\gamma_1+\beta_2\gamma_2-1)\delta(\alpha_1\omega_2+\alpha_2\omega_1-\beta_1\gamma_2-\beta_2\gamma_1)d\omega_1 d\omega_2 D\alpha D\beta D\gamma\\
&=\frac{1}{16\pi^4}\frac{D\alpha D\beta D\gamma}{\left|\alpha\right|^2}=\frac{1}{16\pi^4}\frac{d\alpha_1 d\alpha_2 d\beta_1 d\beta_2 d\gamma_1 d\gamma_2}{\left|\alpha\right|^2}. \label{dg0}
\end{aligned}
\end{equation}
The following calculation can show the details for the third step. For convenience, we define
$$
f_1=Re\left(\alpha\omega-\gamma\beta-1\right)=\alpha_1\omega_1-\alpha_2\omega_2-\beta_1\gamma_1+\beta_2\gamma_2-1,
$$
$$
f_2=Im\left(\alpha\omega-\gamma\beta-1\right)=\alpha_1\omega_2+\alpha_2\omega_1-\beta_1\gamma_2-\beta_2\gamma_1.
$$
Then, the product of delta function can be written as
$$
\delta(f_1)\delta(f_2)=\frac{\delta(\omega_1-\mathring{\omega}_1)\delta(\omega_2-\mathring{\omega}_2)}{\left|\det\frac{\partial(f_1,f_2)}{\partial(\omega_1,\omega_2)}\right|}, \quad \left|\det\frac{\partial(f_1,f_2)}{\partial(\omega_1,\omega_2)}\right|=\alpha_1^2+\alpha_2^2=\left|\alpha\right|^2,
$$
here, $\mathring{\omega}_1$ and $\mathring{\omega}_2$ are the solutions of the system fo equations $f_1=0$ and $f_2=0$,
$$
\mathring{\omega}_1=\frac{\alpha_1+\alpha_1\beta_1\gamma_1+\alpha_2\beta_2\gamma_1+\alpha_2\beta_1\gamma_2-\alpha_1\beta_2\gamma_2}{\alpha_1^2+\alpha_2^2},
$$
$$
\mathring{\omega}_2=\frac{\alpha_1(\beta_2\gamma_1+\beta_1\gamma_2)+\alpha_2(-1-\beta_1\gamma_1+\beta_2\gamma_2}{\alpha_1^2+\alpha_2^2}.
$$
Next, we parametrized 
$$
\alpha=1+\frac{1}{\sqrt{2}}\left(x_1+i y_1\right),\beta=\frac{1}{\sqrt{2}}\left(x_2+i y_2\right),\gamma=\frac{1}{\sqrt{2}}\left(x_3+i y_3\right).
$$
i.e.,
$$
\alpha_1=1+\frac{x_1}{\sqrt{2}},\alpha_2 =\frac{y_1}{\sqrt{2}},\beta_1=\frac{x_2}{\sqrt{2}},\beta_2= \frac{y_2}{\sqrt{2}},\gamma_1=\frac{x_3}{\sqrt{2}},\gamma_2=\frac{y_3}{\sqrt{2}}.
$$
Then, (\ref{dg0}) can be written as 
\begin{equation}
\begin{aligned}
dg&=\frac{1}{16\pi^4\times 2^3}\frac{dx_1 dy_1 dx_2 dy_2 dx_3 dy_3}{\left|1+\frac{x_1+i y_1}{\sqrt{2}}\right|^2}, \label{dg1}
\end{aligned}
\end{equation}
which is the SL$\left(2,\mathbb{C}\right)$ group haar measure we used in our case. The parameters of the SL$\left(2,\mathbb{C}\right)$ group are 
\begin{equation}
g=\left(\begin{matrix}1+\frac{x_1+i y_1}{\sqrt{2}}&\frac{x_2+i y_2}{\sqrt{2}}\\\frac{x_3+i y_3}{\sqrt{2}}&\frac{1+\frac{x_2+i y_2}{\sqrt{2}}\frac{x_3+i y_3}{\sqrt{2}}}{1+\frac{x_1+i y_1}{\sqrt{2}}}\end{matrix}\right).
\end{equation}
At the critical point $g=\ident$ or $\vec{x}=\vec{y}=0$,
\begin{equation}
	dg\rightarrow \frac{1}{16\pi^4\times 2^3}dx_1 dy_1 dx_2 dy_2 dx_3 dy_3.
\end{equation}

\section{Boundary data, critical points and numerical results with different boundary geometries}\label{boundary geometry}

Here we list the boundary data with different boundary geometries $j_{ab}=(8,3)\l$ and $j_{ab}=(11,4)\l$. Table \ref{tab:vertexbd} shows the coordinates of vertices for the boundary geometries $j_{ab}=(8,3)\l$ and $j_{ab}=(11,4)\l$ respectively. Table \ref{tab:4dnormmals} gives the 4-d normal vectors for each tetrahedron for boundary geometries $j_{ab}=(8,3)$ and $j_{ab}=(11,4)$ respectively. Table \ref{tab:xi83} and Table \ref{tab:xi114} gives the boundary state $|\xi_{ab}\rangle$ for each tetrahedron with boundary geometries $j_{ab}=(8,3)\l$ and $j_{ab}=(11,4)\l$ respectively. 
\begin{table}[H]
	\centering\caption{Each cell of the table is the coordinate of the vertex $P_a$ in the Minkowski spacetime.}\label{tab:vertexbd}
	\scriptsize
	\setlength{\tabcolsep}{0.5mm}
	\begin{tabular}{|c|c|c|c|c|c|}
		\hline
		\small\diagbox{}{}&$P_1$&$P_2$&$P_3$&$P_4$&$P_5$\\
		\hline
		$j_{ab}=(8,3)$ &$\left({0, 0, 0, 0}\right)$&$\left(0, 0, 0, -4.298\right)$&$\left(0, 0, -3.722, -2.149\right)$&$\left(0, -3.510, -1.241, -2.149\right)$&$\left(-0.601, -0.8774, -1.241, -2.149\right)$\\
		\hline
		$j_{ab}=(11,4)$ &$\left({0, 0, 0, 0}\right)$&$\left(0, 0, 0, -5.040\right)$&$\left(0, 0, -4.365, -2.520\right)$&$\left(0, -4.115, -1.455, -2.520\right)$&$\left(-0.810, -1.029, -1.455, 
		-2.520\right)$\\
		\hline
	\end{tabular}
\end{table}
\begin{table}[H]
	\centering\caption{Each cell of the table is 4-d normal vectors for each tetrahedron with boundary geometries $j_{ab}=(8,3)\l$ and $j_{ab}=(11,4)\l$ respectively.}\label{tab:4dnormmals}
	\scriptsize
	\setlength{\tabcolsep}{0.5mm}
	\begin{tabular}{|c|c|c|c|c|c|}
		\hline
		\small\diagbox{}{}&$N_1$&$N_2$&$N_3$&$N_4$&$N_5$\\
		\hline
		$j_{ab}=(8,3)$ &$\left(-1, 0, 0, 0\right)$&$\left(1.37, 0.94, 0., 0.\right)$&$\left(1.37, -0.31, 0.89, 0.\right)$&$\left(1.37, -0.31, -0.44, 0.77\right)$&$\left(1.37, -0.31, -0.44, -0.77\right)$\\
		\hline
		$j_{ab}=(11,4)$ &$\left(-1, 0, 0, 0\right)$&$\left(1.62, 1.28, 0., 0.\right)$&$\left(1.62, -0.43, 1.20, 0.\right)$&$\left(1.62, -0.43, -0.60, 1.04\right)$&$\left(1.62, -0.43, -0.60, -1.04\right)$\\
		\hline
	\end{tabular}
\end{table}

\begin{table}[H]
	\centering\caption{Each cell of the table is the boundary state $\ket{\xi_{ab}}$ for the face shared by the line number tetrahedra and the column number tetrahedra with boundary geometry $j_{ab}=(8,3)\l$.}\label{tab:xi83}
	\footnotesize
	\setlength{\tabcolsep}{0.8mm}
	\begin{tabular}{|c|c|c|c|c|c|}
		\hline
		\diagbox{\small{a}}{$\ket{\xi_{ab}}$ }{\small{b}}&1&2&3&4&5\\
		\hline
		1&\diagbox{}{}&(0.71,0.71)&(0.71,-0.24+0.67 i)&(0.95,-0.17-0.25 i)&(0.30,-0.55-0.78 i)\\
		\hline
		2&(0.71,-0.71)&\diagbox{}{}&(0.71,0.63 + 0.32I)&(0.84, 0.53 - 0.14 i)&(0.55, 0.81 - 0.21 i)\\
		\hline
		3&(0.71, 0.24 - 0.67 i)&(0.71, 0.09 + 0.70 i)&\diagbox{}{}&(0.84, -0.31 + 0.46 i)&(0.55, -0.47 + 0.69 i)\\
		\hline
		4&(0.30, 0.55 + 0.78 i)&(0.96, 0.07 - 0.26 i)&(0.96, -0.27 - 0.02 i)&\diagbox{}{}&(0.85, -0.30 - 0.42 i)\\
		\hline
		5&(0.95, 0.17 + 0.25 i)&(0.27, 0.25 - 0.93 i)&(0.27, -0.96-0.07 i)&(0.52, -0.49-0.70 i)&\diagbox{}{}\\
		\hline
	\end{tabular}
\end{table}

\begin{table}[H]
	\centering\caption{Each cell of the table is boundary state $\ket{\xi_{ab}}$ for the face shared by the line number tetrahedra and the column number tetrahedra with boundary geometry $j_{ab}=(11,4)\l$.}\label{tab:xi114}
	\footnotesize
	\setlength{\tabcolsep}{0.8mm}
	\begin{tabular}{|c|c|c|c|c|c|}
		\hline
		\diagbox{\small{a}}{$\ket{\xi_{ab}}$ }{\small{b}}&1&2&3&4&5\\
		\hline
		1&\diagbox{}{}&(0.71,0.71)&(0.71,-0.24+0.67 i)&(0.95,-0.17-0.25 i)&(0.30,-0.55-0.78 i)\\
		\hline
		2&(0.71,-0.71)&\diagbox{}{}&(0.71,0.64 + 0.28I)&(0.82, 0.55 - 0.12 i)&(0.57, 0.80 - 0.17 i)\\
		\hline
		3&(0.71, 0.24 - 0.67 i)&(0.71, 0.05 + 0.71 i)&\diagbox{}{}&(0.82, -0.30 + 0.49 i)&(0.57, -0.43 + 0.70 i)\\
		\hline
		4&(0.30, 0.55 + 0.78 i)&(0.96, 0.04 - 0.26 i)&(0.96, -0.26 - 0.05 i)&\diagbox{}{}&(0.87, -0.28 - 0.40 i)\\
		\hline
		5&(0.95, 0.17 + 0.25 i)&(0.26, 0.14 - 0.96 i)&(0.26, -0.95-0.19 i)&(0.50, -0.50-0.71 i)&\diagbox{}{}\\
		\hline
	\end{tabular}
\end{table} 

We also give critical points with different boundary geometries. Table \ref{tab:ga83} and Table \ref{tab:pos83} are critical points with boundary geometry $j_{ab}=(8,3)\l$, Table \ref{tab:ga114} and Table \ref{tab:pos114} are critical points with boundary geometry $j_{ab}=(11,4)\l$.

\begin{table}[H]
	\centering\caption{Each cell of the table is the critical point of a-th tetrahedron group element $g_a^{0(\pm)}$ with boundary geometry $j_{ab}=(8,3)\l$.}\label{tab:ga83}
	\scriptsize
	\setlength{\tabcolsep}{0.5mm}
	\begin{tabular}{|c|c|c|c|c|c|}
		\hline
		\small{a}&1&2&3&4&5\\
		\hline
		$g_a^{0(+)}$ &$\left(\begin{matrix}
		1&0\\	0&1
		\end{matrix}\right)$&$\left(\begin{matrix}
		0.43 i&1.09 i\\	1.09 i&0.43 i
		\end{matrix}\right)$&$\left(\begin{matrix}
		0.43 i&1.03 - 0.36 i\\	-1.03- 0.36 i&0.43 i
		\end{matrix}\right)$&$\left(\begin{matrix}
		1.32 i&-0.51 - 0.36 i\\	0.51 - 0.36 i&-0.46 i
		\end{matrix}\right)$&$\left(\begin{matrix}
		-0.46 i&-0.51 - 0.36 i\\	0.51 - 0.36 i&1.32 i
		\end{matrix}\right)$\\
		\hline
		$g_a^{0(-)}$ &$\left(\begin{matrix}
		1&0\\	0&1
		\end{matrix}\right)$&$\left(\begin{matrix}
		-0.43 i&1.09 i\\	1.09 i&-0.43 i
		\end{matrix}\right)$&$\left(\begin{matrix}
		-0.43 i&1.03 - 0.36 i\\	-1.03- 0.36 i&0.43 i
		\end{matrix}\right)$&$\left(\begin{matrix}
		0.46 i&-0.51 - 0.36 i\\	0.51 - 0.36 i&-1.32 i
		\end{matrix}\right)$&$\left(\begin{matrix}
		-1.32 i&-0.51 - 0.36 i\\	0.51 - 0.36 i&0.46 i
		\end{matrix}\right)$\\
		\hline
	\end{tabular}
\end{table}

\begin{table}[H]
	\centering\caption{Each cell of the table is the critical point of a-th tetrahedron group element $g_a^{0(\pm)}$ with boundary geometry $j_{ab}=(11,4)\l$.}\label{tab:ga114}
	\scriptsize
	\setlength{\tabcolsep}{0.5mm}
	\begin{tabular}{|c|c|c|c|c|c|}
		\hline
		\small{a}&1&2&3&4&5\\
		\hline
		$g_a^{0(+)}$ &$\left(\begin{matrix}
		1&0\\	0&1
		\end{matrix}\right)$&$\left(\begin{matrix}
		0.55 i&1.14 i\\	1.14 i&0.56 i
		\end{matrix}\right)$&$\left(\begin{matrix}
		0.56 i&1.08 - 0.38 i\\	-1.08- 0.38 i&0.56 i
		\end{matrix}\right)$&$\left(\begin{matrix}
		1.49 i&-0.54 - 0.38 i\\	0.54 - 0.38 i&-0.38 i
		\end{matrix}\right)$&$\left(\begin{matrix}
		-0.38 i&-0.54 - 0.38 i\\	0.54 - 0.38 i&1.49 i
		\end{matrix}\right)$\\
		\hline
		$g_a^{0(-)}$ &$\left(\begin{matrix}
		1&0\\	0&1
		\end{matrix}\right)$&$\left(\begin{matrix}
		-0.56 i&1.44 i\\	1.14 i&-0.56 i
		\end{matrix}\right)$&$\left(\begin{matrix}
		-0.56 i&1.08 - 0.38 i\\	-1.08- 0.38 i&-0.56 i
		\end{matrix}\right)$&$\left(\begin{matrix}
		0.38 i&-0.54 - 0.38 i\\	0.54 - 0.38 i&-1.48 i
		\end{matrix}\right)$&$\left(\begin{matrix}
		-1.49 i&-0.54 - 0.38 i\\	0.54 - 0.38 i&0.38 i
		\end{matrix}\right)$\\
		\hline
	\end{tabular}
\end{table} 

\begin{table}[H]
	\centering\caption{Each cell of the table is the critical point parameterized by $(\theta_{ab}^{0(\pm)}, \phi_{ab}^{0(\pm)})$ for the face shared by the line number tetrahedron a and the column number tetrahedron b, $a<b$. Two tables list the result for two distinct critical points with boundary geometry $j_{ab}=(8,3)\l$.}\label{tab:pos83}
	\footnotesize
	\setlength{\tabcolsep}{0.8mm}
	\begin{tabular}{|c|c|c|c|c|}
		\hline
		\diagbox{\small{a}}{$(\theta_{ab}^{0(+)}, \phi_{ab}^{0(+)})$ }{\small{b}}&2&3&4&5\\
		\hline
		1&(-1.57,0)&(-1.57,1.91)&(-2.53,-2.19)&(-0.62,-2.19)\\
		\hline
		2&\diagbox{}{}&(-1.57,-1.02)&(-0.74, 0.69)&(-2.40, 0.69)\\
		\hline
		3&\diagbox{}{}&\diagbox{}{}&(0.74, 1.23)&(-2.40, 1.23)\\
		\hline
		4&\diagbox{}{}&\diagbox{}{}&\diagbox{}{}&(-2.74, 0.96)\\
		\hline
	\end{tabular}
	\begin{tabular}{|c|c|c|c|c|}
		\hline
		\diagbox{\small{a}}{$(\theta_{ab}^{0(-)}, \phi_{ab}^{0(-)})$ }{\small{b}}&2&3&4&5\\
		\hline
		1&(-1.57,0)&(-1.57,1.91)&(-2.53,-2.19)&(-0.62,-2.19)\\
		\hline
		2&\diagbox{}{}&(-1.57,-0.21)&(-1,39, 0.11)&(-1.75, 0.11)\\
		\hline
		3&\diagbox{}{}&\diagbox{}{}&(-1.39, 1.80)&(-1.75, 1.80)\\
		\hline
		4&\diagbox{}{}&\diagbox{}{}&\diagbox{}{}&(-2.74, -2.19)\\
		\hline
	\end{tabular}
\end{table}

\begin{table}[H]
	\centering\caption{Each cell of the table is critical point parameterized by $(\theta_{ab}^{0(\pm)}, \phi_{ab}^{0(\pm)})$ for the face shared by the line number tetrahedron a and the column number tetrahedron b, $a<b$. Two tables list the result for two distinct critical points with boundary geometry $j_{ab}=(11,4)\l$.}\label{tab:pos114}
	\footnotesize
	\setlength{\tabcolsep}{0.8mm}
	\begin{tabular}{|c|c|c|c|c|}
		\hline
		\diagbox{\small{a}}{$(\theta_{ab}^{0(+)}, \phi_{ab}^{0(+)})$ }{\small{b}}&2&3&4&5\\
		\hline
		1&(-1.57,0)&(-1.57,1.91)&(-2.53,-2.19)&(-0.62,-2.19)\\
		\hline
		2&\diagbox{}{}&(-1.57,-1.09)&(-0.70, 0.76)&(-2.44, 0.76)\\
		\hline
		3&\diagbox{}{}&\diagbox{}{}&(-0.70, 1.15)&(-2.44, 1.15)\\
		\hline
		4&\diagbox{}{}&\diagbox{}{}&\diagbox{}{}&(-2.67, 0.96)\\
		\hline
	\end{tabular}
	\begin{tabular}{|c|c|c|c|c|}
		\hline
		\diagbox{\small{a}}{$(\theta_{ab}^{0(-)}, \phi_{ab}^{0(-)})$ }{\small{b}}&2&3&4&5\\
		\hline
		1&(-1.57,0)&(-1.57,1.91)&(-2.53,-2.19)&(-0.62,-2.19)\\
		\hline
		2&\diagbox{}{}&(-1.57,-0.14)&(-1.45, 0.07)&(-1.70, 0.07)\\
		\hline
		3&\diagbox{}{}&\diagbox{}{}&(-1.45, 1.84)&(-1.70, 1.84)\\
		\hline
		4&\diagbox{}{}&\diagbox{}{}&\diagbox{}{}&(-2.67, -2.19)\\
		\hline
	\end{tabular}
\end{table}

With these boundary data and critical points for the different boundary geometries, we can compute the asymptotics of the EPRL 4-simplex amplitude with the next-to-leading-order corrections. We take $\g=0.1$ as an example for the following results. For the coherent intertwiner as the boundary state, the asymptotic amplitude $A^{(8,3)}_v$ with boundary geometry $j_{ab}=(8,3)\l$ can be written as:   
\be
A^{(8,3)}_v&=&\left(1+\frac{1}{4 \lambda }\right)^6\left(1+\frac{1}{10 \lambda }\right)^4\frac{6.41 \times 10^{-18}}{\lambda^{12}}\,e^{10.71 \lambda i}\nonumber\\
&&\left[\cos (0.18+0.14 \lambda)+\frac{1.71}{\lambda} \sin (-1.28+0.14 \lambda)+O\lt(\frac{1}{\l^2}\rt)\right].\label{Av3and8}
\ee 
The asymptotic amplitude $A^{(11,4)}_v$ with boundary geometry $j_{ab}=(11,4)\l$ is
\be
A^{(11,4)}_v&=&\left(1+\frac{1}{4 \lambda }\right)^6\left(1+\frac{1}{10 \lambda }\right)^4\frac{5.59 \times 10^{-21}}{\lambda^{12}}\,e^{4.26 \lambda i}\nonumber\\
&&\left[\cos (0.19+0.31 \lambda)+\frac{1.21}{\lambda} \sin (-1.24+0.31 \lambda)+O\lt(\frac{1}{\l^2}\rt)\right].\label{Av4and11}
\ee 
For the coherent spin-network as the boundary state, the asymptotic amplitude $A'^{(8,3)}_v$ with the boundary geometry $j_{ab}=(8,3)\l$ is 
{\be
	A'^{(8,3)}_v &=&2^{10}\,\left(1+\frac{1}{4 \lambda }\right)^6\left(1+\frac{1}{10 \lambda }\right)^4\frac{8.94\times 10^{-12}+3.85\times 10^{-12} i}{\lambda^{7}}\,\e^{10.85 \lambda i }\nonumber\\
	&&\left[1+\frac{2.06+0.88 i}{\lambda}+O\left(\frac{1}{\l^2}\right)\right],
	\label{coh3and8}
	\ee}
and the asymptotic amplitude $A'^{(11,4)}_v$ with the boundary geometry $j_{ab}=(11,4)\l$ is 
{\be
	A'^{(11,4)}_v &=&2^{10}\,\left(1+\frac{1}{4 \lambda }\right)^6\left(1+\frac{1}{10 \lambda }\right)^4\frac{3.70\times 10^{-13}+1.67\times 10^{-13} i}{\lambda^{7}}\,\e^{4.57 \lambda i }\nonumber\\
	&&\left[1+\frac{-0.43-0.25 i}{\lambda}+O\left(\frac{1}{\l^2}\right)\right].
	\label{coh4and11}
	\ee}

\bibliographystyle{jhep}
\bibliography{note}

\providecommand{\href}[2]{#2}\begingroup\raggedright\begin{thebibliography}{10}

\bibitem{Thiemann:2007pyv}
T.~Thiemann, {\em Modern Canonical Quantum General Relativity}.
\newblock Cambridge Monographs on Mathematical Physics. Cambridge University
  Press, Cambridge, England, 2007.

\bibitem{review1}
A.~Ashtekar and J.~Lewandowski, {\it {Background independent quantum gravity: A
  Status report}},  {\em Class.Quant.Grav.} {\bf 21} (2004) R53,
  [\href{http://arxiv.org/abs/gr-qc/0404018}{{\tt gr-qc/0404018}}].

\bibitem{Han:2005km}
M.~Han, W.~Huang, and Y.~Ma, {\it {Fundamental structure of loop quantum
  gravity}},  {\em Int. J. Mod. Phys. D} {\bf 16} (2007) 1397--1474,
  [\href{http://arxiv.org/abs/gr-qc/0509064}{{\tt gr-qc/0509064}}].

\bibitem{rovelli2014covariant}
C.~Rovelli and F.~Vidotto, {\em Covariant Loop Quantum Gravity: An Elementary
  Introduction to Quantum Gravity and Spinfoam Theory}.
\newblock Cambridge Monographs on Mathematical Physics. Cambridge University
  Press, 2014.

\bibitem{Perez2012}
A.~Perez, {\it {The Spin Foam Approach to Quantum Gravity}},  {\em Living
  Rev.Rel.} {\bf 16} (2013) 3, [\href{http://arxiv.org/abs/1205.2019}{{\tt
  arXiv:1205.2019}}].

\bibitem{Rovelli:2010vv}
C.~Rovelli, {\it {Simple model for quantum general relativity from loop quantum
  gravity}},  {\em J. Phys. Conf. Ser.} {\bf 314} (2011) 012006,
  [\href{http://arxiv.org/abs/1010.1939}{{\tt arXiv:1010.1939}}].

\bibitem{Barrett:2009mw}
J.~W. Barrett, R.~J. Dowdall, W.~J. Fairbairn, F.~Hellmann, and R.~Pereira,
  {\it {Lorentzian spin foam amplitudes: Graphical calculus and asymptotics}},
  {\em Class. Quant. Grav.} {\bf 27} (2010) 165009,
  [\href{http://arxiv.org/abs/0907.2440}{{\tt arXiv:0907.2440}}].

\bibitem{Han:2011re}
M.~Han and M.~Zhang, {\it {Asymptotics of Spinfoam Amplitude on Simplicial
  Manifold: Lorentzian Theory}},  {\em Class. Quant. Grav.} {\bf 30} (2013)
  165012, [\href{http://arxiv.org/abs/1109.0499}{{\tt arXiv:1109.0499}}].

\bibitem{Engle:2007wy}
J.~Engle, E.~Livine, R.~Pereira, and C.~Rovelli, {\it {LQG vertex with finite
  Immirzi parameter}},  {\em Nucl. Phys. B} {\bf 799} (2008) 136--149,
  [\href{http://arxiv.org/abs/0711.0146}{{\tt arXiv:0711.0146}}].

\bibitem{Conrady:2008ea}
F.~Conrady and L.~Freidel, {\it {Path integral representation of spin foam
  models of 4d gravity}},  {\em Class. Quant. Grav.} {\bf 25} (2008) 245010,
  [\href{http://arxiv.org/abs/0806.4640}{{\tt arXiv:0806.4640}}].

\bibitem{Han:2013gna}
M.~Han and T.~Krajewski, {\it {Path Integral Representation of Lorentzian
  Spinfoam Model, Asymptotics, and Simplicial Geometries}},  {\em Class. Quant.
  Grav.} {\bf 31} (2014) 015009, [\href{http://arxiv.org/abs/1304.5626}{{\tt
  arXiv:1304.5626}}].

\bibitem{Dona:2019dkf}
P.~Dona, M.~Fanizza, G.~Sarno, and S.~Speziale, {\it {Numerical study of the
  Lorentzian Engle-Pereira-Rovelli-Livine spin foam amplitude}},  {\em Phys.
  Rev.} {\bf D100} (2019), no.~10 106003,
  [\href{http://arxiv.org/abs/1903.12624}{{\tt arXiv:1903.12624}}].

\bibitem{Hormander}
L.~Hormander, {\em The Analysis of Linear Partial Differential Operators I}.
\newblock Springer-Verlag Berlin, 1983.

\bibitem{Bonzom:2008xd}
V.~Bonzom, E.~R. Livine, M.~Smerlak, and S.~Speziale, {\it {Towards the
  graviton from spinfoams: The Complete perturbative expansion of the 3d toy
  model}},  {\em Nucl. Phys. B} {\bf 804} (2008) 507--526,
  [\href{http://arxiv.org/abs/0802.3983}{{\tt arXiv:0802.3983}}].

\bibitem{qudx.org}
M.~Han, Z.~Huang, H.~Liu, and D.~Qu.
  \url{https://github.com/dqu2017/Numerical-Asymtotics}, 2020.

\bibitem{Dona:2018nev}
P.~Dona and G.~Sarno, {\it {Numerical methods for EPRL spin foam transition
  amplitudes and Lorentzian recoupling theory}},  {\em Gen. Rel. Grav.} {\bf
  50} (2018) 127, [\href{http://arxiv.org/abs/1807.03066}{{\tt
  arXiv:1807.03066}}].

\bibitem{Bahr:2016hwc}
B.~Bahr and S.~Steinhaus, {\it {Numerical evidence for a phase transition in 4d
  spin foam quantum gravity}},  {\em Phys. Rev. Lett.} {\bf 117} (2016), no.~14
  141302, [\href{http://arxiv.org/abs/1605.07649}{{\tt arXiv:1605.07649}}].

\bibitem{Bahr:2018gwf}
B.~Bahr, G.~Rabuffo, and S.~Steinhaus, {\it {Renormalization of symmetry
  restricted spin foam models with curvature in the asymptotic regime}},  {\em
  Phys. Rev. D} {\bf 98} (2018), no.~10 106026,
  [\href{http://arxiv.org/abs/1804.00023}{{\tt arXiv:1804.00023}}].

\bibitem{Conrady:2008mk}
F.~Conrady and L.~Freidel, {\it {On the semiclassical limit of 4d spin foam
  models}},  {\em Phys. Rev.} {\bf D78} (2008) 104023,
  [\href{http://arxiv.org/abs/0809.2280}{{\tt arXiv:0809.2280}}].

\bibitem{Barrett:2009gg}
J.~W. Barrett, R.~J. Dowdall, W.~J. Fairbairn, H.~Gomes, and F.~Hellmann, {\it
  {Asymptotic analysis of the EPRL four-simplex amplitude}},  {\em J. Math.
  Phys.} {\bf 50} (2009) 112504, [\href{http://arxiv.org/abs/0902.1170}{{\tt
  arXiv:0902.1170}}].

\bibitem{Han:2011rf}
M.-X. Han and M.~Zhang, {\it {Asymptotics of Spinfoam Amplitude on Simplicial
  Manifold: Euclidean Theory}},  {\em Class. Quant. Grav.} {\bf 29} (2012)
  165004, [\href{http://arxiv.org/abs/1109.0500}{{\tt arXiv:1109.0500}}].

\bibitem{Freidel:2007py}
L.~Freidel and K.~Krasnov, {\it {A New Spin Foam Model for 4d Gravity}},  {\em
  Class. Quant. Grav.} {\bf 25} (2008) 125018,
  [\href{http://arxiv.org/abs/0708.1595}{{\tt arXiv:0708.1595}}].

\bibitem{Peskin:257493}
M.~E. Peskin and D.~V. Schroeder, {\em {An introduction to quantum field
  theory}}.
\newblock Westview, Boulder, CO, 1995.
\newblock Includes exercises.

\bibitem{Rovelli:2005yj}
C.~Rovelli, {\it {Graviton propagator from background-independent quantum
  gravity}},  {\em Phys. Rev. Lett.} {\bf 97} (2006) 151301,
  [\href{http://arxiv.org/abs/gr-qc/0508124}{{\tt gr-qc/0508124}}].

\bibitem{Bianchi:2009ky}
E.~Bianchi, E.~Magliaro, and C.~Perini, {\it {Coherent spin-networks}},  {\em
  Phys. Rev.} {\bf D82} (2010) 024012,
  [\href{http://arxiv.org/abs/0912.4054}{{\tt arXiv:0912.4054}}].

\bibitem{Thiemann:2000bw}
T.~Thiemann, {\it {Gauge field theory coherent states (GCS): 1. General
  properties}},  {\em Class. Quant. Grav.} {\bf 18} (2001) 2025--2064,
  [\href{http://arxiv.org/abs/hep-th/0005233}{{\tt hep-th/0005233}}].

\bibitem{Bianchi:2010mw}
E.~Bianchi, E.~Magliaro, and C.~Perini, {\it {Spinfoams in the holomorphic
  representation}},  {\em Phys. Rev. D} {\bf 82} (2010) 124031,
  [\href{http://arxiv.org/abs/1004.4550}{{\tt arXiv:1004.4550}}].

\bibitem{Dona:2020tvv}
P.~Dona, F.~Gozzini, and G.~Sarno, {\it {Numerical analysis of spin foam
  dynamics and the flatness problem}},
  \href{http://arxiv.org/abs/2004.12911}{{\tt arXiv:2004.12911}}.

\bibitem{Witten:2010cx}
E.~Witten, {\it {Analytic Continuation Of Chern-Simons Theory}},  {\em AMS/IP
  Stud. Adv. Math.} {\bf 50} (2011) 347--446,
  [\href{http://arxiv.org/abs/1001.2933}{{\tt arXiv:1001.2933}}].

\bibitem{Bedaque:2017epw}
P.~F. Bedaque, {\it {A complex path around the sign problem}},  {\em EPJ Web
  Conf.} {\bf 175} (2018) 01020, [\href{http://arxiv.org/abs/1711.05868}{{\tt
  arXiv:1711.05868}}].

\bibitem{toappear}
M.~Han, Z.~Huang, H.~Liu, and D.~Qu, {\it {Monte-Carlo simulation of spinfoam
  amplitude on Lefschetz thimble}},  {\em to appear}.

\bibitem{Ruhl}
W.~Ruhl, {\em The Lorentz Group and Harmonic Analysis}.
\newblock W.A. Benjamin, New York, 1970.

\end{thebibliography}\endgroup
\end{document}